%% file: allocation.tex
\pdfoutput=1
\def\year{2022}\relax
\documentclass[letterpaper]{article} 
\usepackage{aaai22}  
\usepackage{times}  
\usepackage{helvet}  
\usepackage{courier}  
\usepackage[hyphens]{url}  
\usepackage{graphicx} 
\usepackage{natbib}  
\usepackage{caption} 
\DeclareCaptionStyle{ruled}{labelfont=normalfont,labelsep=colon,strut=off} 
\nocopyright
\title{Allocation Schemes in Analytic Evaluation:\\Applicant-Centric Holistic or Attribute-Centric Segmented?}
\author{
    Jingyan Wang,\textsuperscript{\rm 1,2}
    Carmel Baharav,\textsuperscript{\rm 1}
    Nihar B. Shah,\textsuperscript{\rm 1}
    Anita Williams Woolley,\textsuperscript{\rm 1}
    R Ravi\textsuperscript{\rm 1}
}
\affiliations{
    \textsuperscript{\rm 1} Carnegie Mellon University\\
    \textsuperscript{\rm 2} Georgia Institute of Technology
}

\usepackage{xspace,enumitem,subfig,mathtools,dsfont,amsthm,amssymb,amsfonts,amsmath,cleveref}
\newtheorem{hypothesis}{Hypothesis}

\input{standard_math}
\input{macros}

\newcommand{\hyphen}{\text{-}} 
\newcommand{\figwidth}{0.69\linewidth}
\begin{document}

\maketitle

\begin{abstract}
\input{text_abstract}
\end{abstract}

\input{text_main}

\bibliography{references}

\onecolumn
\appendix
\input{text_appendix}

\end{document}

%% file: standard_math.tex
\newcommand{\Prob}{\mathbb{P}}

\newcommand{\Probbig}[1]{\Prob\Big(#1\Big)}
\newcommand{\sample}{\sim}
\newcommand{\defn}{\colon=}
\newcommand{\given}{\mid}

\newtheorem{theorem}{Theorem}

\newcommand{\setcomplement}[1]{\overline{#1}}

\newcommand{\union}{\cup}
\newcommand{\normal}{\mathcal{N}}

\newcommand{\reals}{\mathbb{R}}

\newcommand{\vecone}{\mathbf{1}}

\DeclarePairedDelimiter\abs{\lvert}{\rvert}

\DeclareMathOperator*{\argmax}{argmax}

\newcommand{\stepone}{\text{(i)}\xspace}
\newcommand{\steptwo}{\text{(ii)}\xspace}

\newcommand{\textth}{\textrm{th}}

\newcommand*{\approxident}{%
  \mathrel{\vcenter{\offinterlineskip
  \hbox{$\sim$}\vskip-.35ex\hbox{$\sim$}\vskip-.35ex\hbox{$\sim$}}}}
\def\apeq{\approxident}

%% file: macros.tex
\newcommand{\applicant}{applicant\xspace}
\newcommand{\applicants}{applicants\xspace}
\newcommand{\reviewer}{evaluator\xspace}
\newcommand{\reviewers}{evaluators\xspace}

\newcommand{\attribute}{attribute\xspace}
\newcommand{\attributes}{attributes\xspace}

\newcommand{\holistic}{holistic\xspace}
\newcommand{\Holistic}{Holistic\xspace}
\newcommand{\grouptwenty}{$20$Q-group\xspace}
\newcommand{\groupfive}{$5$Q-group\xspace}
\newcommand{\grouptwentyinit}{20Q-initial\xspace} 

\newcommand{\segmented}{segmented\xspace}
\newcommand{\Segmented}{Segmented\xspace}
\newcommand{\segallo}{\segmented allocation\xspace}
\newcommand{\Segallo}{\Segmented allocation\xspace}
\newcommand{\holallo}{\holistic allocation\xspace}

\newcommand{\numapps}{n}
\newcommand{\numattrs}{d}
\newcommand{\idxapp}{i}
\newcommand{\idxattr}{j}
\newcommand{\idxpair}{{\idxapp\idxattr}}
\newcommand{\distval}{\mathcal{D}}

\newcommand{\event}{E}
\newcommand{\eventerror}{E}
\newcommand{\eventerrorhol}{\eventerror_\texthol}
\newcommand{\eventerrorseg}{\eventerror_\textseg}

\newcommand{\eventbiasedreviewer}{R}
\newcommand{\err}{e}
\newcommand{\textseg}{\textrm{seg}}
\newcommand{\texthol}{\textrm{hol}}
\newcommand{\errseg}{\err_\textseg}
\newcommand{\errhol}{\err_\texthol}

\newcommand{\parampowerlaw}{\delta}
\newcommand{\fracbiased}{\gamma}
\newcommand{\fracappsdisadvantage}{\alpha}
\newcommand{\discount}{\beta}
\newcommand{\fracfeatsdisadvantage}{\lambda}
\newcommand{\valdisadvantage}{x}
\newcommand{\valadvantage}{x}

\newcommand{\textmax}{\textrm{max}}
\newcommand{\textdisadvantage}{\textrm{d}}
\newcommand{\textadvantage}{\textrm{a}}
\newcommand{\valmaxdisadvantage}{\valdisadvantage^{\textmax}_\textdisadvantage}
\newcommand{\valmaxadvantage}{\valadvantage^{\textmax}_\textadvantage}

\newcommand{\setapps}{S}
\newcommand{\setdisadvantage}{\setapps_\textdisadvantage}
\newcommand{\setadvantage}{\setapps_\textadvantage}
\newcommand{\funcbin}{\omega} 

\newcommand{\cellmtx}{x}
\newcommand{\vecpercentiles}{\boldsymbol{z}}
\newcommand{\score}{y}

\newcommand{\corr}{\sigma}
\newcommand{\thresh}{\tau}

%% file: text_abstract.tex
Many applications such as hiring and university admissions involve evaluation and selection of applicants. These tasks are fundamentally difficult, and require combining evidence from multiple different aspects (what we term ``attributes''). In these applications, the number of applicants is often large, and a common practice is to assign the task to multiple \reviewers in a distributed fashion. Specifically, in the often-used \textbf{\holistic} allocation, each \reviewer is assigned a subset of the applicants, and is asked to assess all relevant information for their assigned applicants. However, such an evaluation process is subject to issues such as miscalibration (\reviewers see only a small fraction of the applicants and may not get a good sense of relative quality), and discrimination (\reviewers are influenced by irrelevant information about the applicants). We identify that such attribute-based evaluation allows alternative allocation schemes. Specifically, we consider assigning each \reviewer more applicants but fewer attributes per applicant, termed \textbf{segmented} allocation. We compare \segmented allocation to \holistic allocation on several dimensions via theoretical and experimental methods. We establish various tradeoffs between these two approaches, and identify conditions under which one approach results in more accurate evaluation than the other.

%% file: text_main.tex
\section{Introduction}

Evaluation and selection are two essential functions that play a critical role in almost every organization as they determine who joins the organization, who remains, and the resulting organizational performance. However, they can also be a significant source of errors, leading to bad selection decisions and potentially limiting opportunities for certain groups. In the past, concerns about inaccurate or biased selection decisions have led to recommendations for the use of structured processes, such as structured job interviews, so that they are consistently conducted and fair to all \applicants~\citep{schmidt1998validity}.

\begin{figure*}[t]
    \centering
    \subfloat[]{
        \begin{minipage}[t]{0.09\linewidth}\raisebox{0.62cm}{\includegraphics[width=\linewidth]{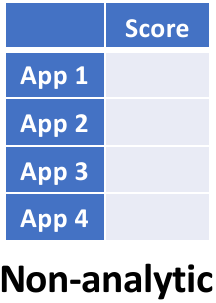}}\end{minipage}
        \hspace{0.8cm}\vrule\hspace{0.8cm}
        \includegraphics[width=0.606\linewidth]{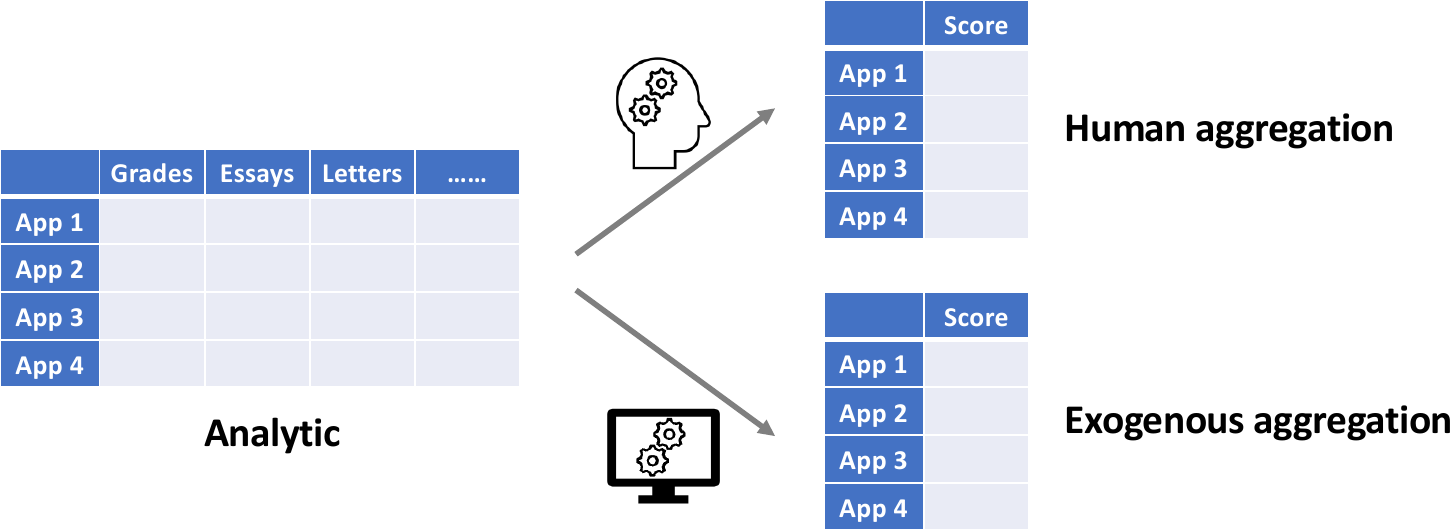}
        \label{float:scheme}
    }
    
  \hrule

  \subfloat[]{\includegraphics[width=0.729\linewidth]{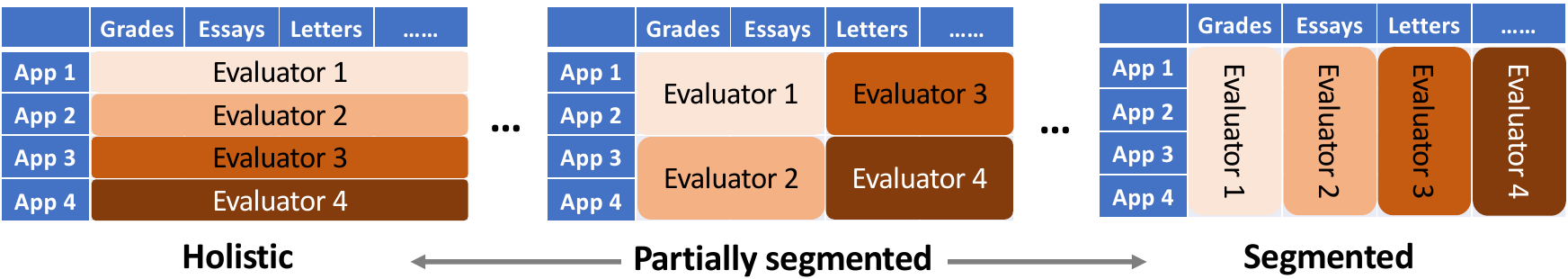}\label{float:allocation}}

  \caption{
    An illustration of the difference between non-analytic and analytic evaluation approaches (top panel), and the spectrum of holistic vs. segmented allocation (bottom panel).\label{fig:scheme}
  }
\end{figure*}

At the other end of the spectrum involving lower-stakes selection problems, distribution and automation of the evaluation task has become popular.
Over the last few decades, developments in online collaboration and decision making have demonstrated the benefits of using the ``crowd'' for many decisions \cite{surowiecki2005wisdom}, opening up new possibilities for conducting evaluation and selection in a more efficient, accurate, and potentially less biased manner. For some decisions, crowd-based processes produce better results when aggregating human inputs by algorithms. However, there have also been ample examples of less effective decisions arrived at by crowds \cite{hube2019understanding}. In reviewing the typical approaches to these crowd-based evaluations and decisions, it appears that the way they are structured varies considerably in terms of the kinds of information reviewed by evaluators when making decisions \cite{draws2021checklist}. We investigate the intricacies when taking this idea of distributed judgment back to the high-stakes regime, and study how the structure of an evaluation process influences the quality of decisions. 

Figure~\ref{fig:scheme}\subref{float:scheme} summarizes the design choices involved in an evaluation procedure. 
In this work, we focus on \textbf{\textit{analytic}} evaluation, where the evaluation of an \applicant (e.g., a job candidate) is decomposed into a pre-defined set of attributes. Analytic evaluation is commonly used in hiring, admissions and grading. For example, in admissions, the attributes may include the student's school GPA, essay quality, and the strength of recommendations letters. On the other hand, in \textbf{\textit{non-analytic}} evaluation, the \reviewer is not required to separately examine individual attributes. Instead, it is sufficient to provide an overall score to each \applicant. While not defining attributes in the non-analytic regime offers \reviewers the freedom to comprehensively think about all possible aspects of the \applicants, the lack of structure  may cause the \reviewers to overly rely on their general impression, leading to inconsistency and inaccuracy as compared to the analytic approach~\cite{jonsson2021analytic}. Hence, analytic evaluation is our regime of interest.

Another design choice is the method to aggregate attributes to derive an overall score for each \applicant. We consider \textbf{\textit{exogenous}} aggregation, where attributes are aggregated using pre-defined (the simplest example is to take a mean, or a weighted mean, of all attributes), or algorithmically learned~\citep{noothigattu2018choosing} rules. On the other hand, \textbf{\textit{human}} aggregation means that after evaluating individual attributes, the \reviewer additionally provides a final score by combining the attributes in some sensible way of the \reviewer's choice. Although human aggregation hypothetically provides more flexibility, simple exogenous aggregation rules often turn out to be no less accurate, or even outperform human aggregation~\citep{kahneman2021noise}. All in all, the issues with the non-analytic approach or human aggregation reveal that the supremacy of human reasoning is overestimated: ``People trust that the complex characteristics of applicants can be best assessed by a sensitive, equally complex human being. This does not stand up to scientific scrutiny''~\citep{highhouse2008stubborn}. Hence, our regime of interest is \textbf{analytic evaluation under exogenous aggregation}.

A fundamental question in structuring the evaluation process is how to allocate \applicants and attributes to \reviewers. There are two basic approaches which are likely to lead to different outcomes (see Figure~\ref{fig:scheme}\subref{float:allocation}). First, in \textbf{\textit{\holistic}} allocation, \reviewers are asked to review and assess all attributes about each \applicant. As shown in Figure~\ref{fig:scheme}\subref{float:allocation}, if we assume each \reviewer is represented by a rectangle of a fixed area (workload), then \holistic allocation entails rectangles of the longest width (number of attributes), and therefore the smallest height (number of \applicants). In \textbf{\textit{segmented}} allocation, if we hold the workload constant, each \reviewer reviews one or a few attributes for a larger number of \applicants (see the right end of Figure~\ref{fig:scheme}\subref{float:allocation}).

\Holistic allocation is quite common in organizational hiring processes as well as in academic admissions~\citep{de2021revising}, where people feel that a more complete understanding of an applicant or the nuances of human judgment improves the quality of decisions. \Segmented allocation is more likely to be used when the attributes are considered relatively independent of one another. One example where a segmented approach is common is the grading of assignments in educational settings, where different instructors may grade different questions since performance on one is not viewed as relevant to the evaluation of another. 

Here, however, we raise the question of \textit{whether holistic allocation is as effective as often assumed, or might it be the case that \segmented allocation would result in better decisions}? Certainly, for \segmented allocation, it is necessary that the attributes being evaluated are separable enough that independent evaluations of them are feasible, which holds true in many instances. In these instances, we argue that \segmented allocation could result in better decision quality, at least under certain conditions. 

We provide a brief review of the relevant literature and outline a framework describing the key difficulties associated with evaluation that have been identified in extant research, including \textbf{calibration} of evaluators, \textbf{efficiency} with which evaluation is conducted, and the degree of \textbf{bias} in the resulting decisions. In presenting our framework, we also delineate specific conditions under which we expect that holistic or segmented allocation leads to better decision outcomes. 

We employ a mixed-method study combining modeling, simulation, crowdsourcing experiments and theoretical analysis to explore the conditions under which \holistic or segmented allocation performs better in terms of calibration, efficiency and fairness. In brief, we find that \segmented allocation provides benefits for evaluator's calibration accuracy, whereas \holistic allocation leads to greater efficiency in carrying out evaluations. In terms of mitigating bias, we observe mixed results depending on specific conditions of the application under consideration. Taken together, our work integrates a few lines of research with implications for the quality of evaluation decisions in a variety of different environments, and provides guidance to system designers for determining the evaluation structure that works best in their context.

We discuss a few key differences between our work and prior literature. At a high level, \holistic and \segmented allocations concern about decomposing and distributing a complex task into smaller parts. In crowdsourcing, complex tasks, such as creating animated movies, making course videos or writing articles, are broken down into parts in a similar fashion. Different crowdsourcing workers complete different parts, and then their work is computationally or manually put together~\cite{kittur2011forge,retelny2014expert,cheng2015break}. While these works measure the quality of the completed work, such as by rating the written articles by professional journalists, we focus on more concrete and quantitative impacts of such decomposition in an evaluation and selection context, where some considerations we study such as fairness naturally arise. Another important application of human computation is peer review, where there is a large body of work on assigning reviewers to papers~\cite[Chapters 3 and 4]{shah2022surveyextended}, with a focus on finding the assignments that maximize the expertise of reviewers assigned to the papers or mitigating undesirable behavior by reviewers. In our work, we primarily consider applications where the work for evaluating individual attributes can be effectively decomposed. On the other hand, in peer review, the task usually cannot be readily decomposed, and reviewers are required to read the entire paper.
We discuss more related work when we formally introduce specific dimensions in Section~\ref{sec:background}. 

All experiments conducted in this paper were approved by the Institutional Review Board (IRB) at Carnegie Mellon University.
The crowdsourcing data, the user interface, as well as all code to reproduce our results is available at\\\url{https://github.com/jingyanw/segmented-vs-holistic}.

\section{Theoretical Background} \label{sec:background}
In theorizing the conditions under which \holistic or \segmented allocation leads to better decision making, we identify a few key difficulties from extant research, including \textit{calibration} of evaluators, \textit{efficiency} with which evaluation is conducted, and the degree of \textit{bias} mitigation. Along these key dimensions, we present six hypotheses, study three of them in detail using theory, simulation and experiments, and leave the remaining three hypotheses for future work.

\subsection{Calibration}
In the context of evaluation, we use ``calibration'' to refer to the ability of evaluators to apply consistent criteria in assessing \applicants, such that the evaluation accurately reflects each \applicant's quality relative to the entire pool~\cite{osborne1991statistical}. 
Note that if an evaluator is able to perfectly identify the placement of each \applicant with respect to all others under consideration, then the \reviewer identifies a perfect ranking of all \applicants. However, the following reasons hinder the \reviewer's ability.

\subsubsection{Lack of information about the population.}
In many situations, evaluators lack complete information about the full range of quality represented by \applicants in the pool, and thus are not able to calibrate their assessment perfectly.
Eliciting ordinal data such as pairwise comparisons or rankings~\cite{shah2017design} helps mitigate miscalibration. Nevertheless, ratings have their own benefits~\cite{wang2018your} and ratings of some form are widely used in practice to compare \applicants assessed by different \reviewers. For instance, the \applicants are placed in categories such as \{definitely admit, maybe admit, waitlist, do not admit\} in admissions, and employees are placed in categories such as \{above average, below average\} in performance evaluation~\cite{goffin2011relative}.

We expect that issues related to \reviewer calibration are among the major drawbacks of \holistic allocation. In \holistic  allocation, each \reviewer assesses all attributes for each \applicant they are assigned. With the exception of very small applicant pools, this necessitates that each \reviewer only sees a small subset of the entire pool. By contrast, in \segmented allocation, each \reviewer sees a much larger set of the pool, perhaps even the entire set of scores in the pool for their assigned attributes. Therefore, we expect that \segmented allocation has an advantage of enhancing \reviewer calibration.

Although it seems intuitive that evaluating more \applicants improves calibration, it is unclear if it actually manifests in practice. To see a counter-argument, consider the following pair of scenarios. In the first scenario, the \reviewer reviews $5$ \applicants whereas in the second scenario, the \reviewer reviews $20$ \applicants. One may expect that when evaluating the last few \applicants in the second scenario, the \reviewer has already seen many more applications than in the first scenario. However, the \reviewer may only be able to keep in mind $5$ or fewer \applicants when evaluating any other \applicant, in which case their calibration in both scenarios will be comparable.

\begin{hypothesis}[Studied in Section~\ref{sec:calibration}]\label{h:calibration}
    For each individual attribute, \segmented allocation, in which each \reviewer has access to more \applicants, leads to better calibration.
    \label{hypothesis-accuracy}
\end{hypothesis}

\subsubsection{Ordering effect.}
The lack of information about the population suggests that calibration depends on the total number of \applicants assigned to \reviewers. Calibration may further vary as a function of the ordering that these \applicants are evaluated. One reason for such variation is the bounded rationality of people, such as the cognitive effects of primacy and recency~\cite{page2010idol}, assimilation and contrast~\cite{damisch2006olympic}, and generosity-erosion~\cite{vives2021erosion}. A second reason for such variation is that \reviewers gradually adapt their calibration as they evaluate each \applicant along the way: When an \reviewer rates the $5^\textth$ \applicant, their grading scale is based on the first $5$ \applicants seen so far, but by the time the \reviewer moves to rate the $100^\textth$ \applicant, they have acquired much more information for calibration from the $100$ \applicants compared to when they rate the $5^\textth$ \applicant.

Such ordering effect can be mitigated in \segmented allocation: Since the attributes of the \applicants are assigned to different \reviewers, the ordering can be shuffled so that each \reviewer sees a different ordering, thereby ``averaging out'' the effect of ordering when the scores from these \reviewers are aggregated.

\begin{hypothesis}
   \Segmented allocation, in which the ordering of the applicants can be shuffled independently for each attribute, leads to better calibration compared to \holistic allocation, under which all attributes are evaluated under one ordering by design.\label{hypothesis:ordering}
\end{hypothesis}

\subsection{Efficiency}

Selection and evaluation processes can also be resource-intensive and time-consuming. One reason of why quality might suffer is the basic human tendency to ``satisfice'' \cite{hilbert2012toward}, particularly when workload is high. Consequently, we contend that another important element to consider in evaluating the relative benefits of different allocation schemes is the impact on efficiency. This pertains to how quickly \reviewers make their assessments, but also the degree to which an allocation scheme affords \reviewers an opportunity to find shortcuts to adaptively allocate their effort. 

\subsubsection{Adaptively allocating effort.} The goal of many evaluation and selection processes is to identify the best subset of \applicants from the available pool. In \holistic allocation, if a particular \applicant is clearly below the threshold on a subset of the attributes, the \reviewer may conclude that the \applicant will not be selected, without scrutinizing the remaining attributes or giving a precise score to the \applicant. In addition, evaluators may use signals, such as red flags in recommendation letters in academic admissions, to draw a preliminary conclusion which they quickly confirm or deny with a cursory review of the remaining information.  The evaluators also enjoy the flexibility to adaptively choose which attribute to review next based on the attributes already reviewed.

In contrast, adaptive strategies are more challenging to implement in \segmented allocation, because the evaluation task is typically allocated in parallel to the \reviewers. That said, within \segmented allocation, the system could employ a filtering rule for certain attributes \emph{before} assigning \applicants to \reviewers. For example, in academic admissions, threshold values for standardized test scores and GPAs are often used as preliminary filters to eliminate some \applicants from further consideration. However, there are concerns such that standardized test scores are themselves biased against certain groups of applicants.
Another remedy is to decompose the evaluation task into multiple rounds, where applicants are filtered in between rounds. However, having multiple rounds adds logistical complexity to the evaluation procedure, and  may also require more time to complete the evaluation process.

We hypothesize that in \holistic allocation, evaluators can reap the adaptive benefits of efficiency without significantly sacrificing accuracy. Furthermore, we postulate that the gain is more prominent when the attributes being evaluated are correlated with one another: Screening \applicants primarily based on the assessment of one attribute will be less likely to lead to errors in the overall assessment, when attributes are highly correlated than when they are only weakly correlated or independent.

\begin{hypothesis}[Studied in Section~\ref{sec:efficiency}]\label{h:efficiency}
    \Holistic allocation results in more efficiency in evaluation without significantly reducing accuracy, when the attributes being assessed are highly correlated and thus can be used as proxies or screening tools for one another.
    \label{hypothesis:efficiency}
\end{hypothesis}
   
\subsubsection{Switching costs.}
In \holistic allocation, the \reviewer primarily switches between different \attributes, whereas in \segmented allocation, the \reviewer primarily switches between \applicants. Whether switching between \applicants or \attributes involves greater effort depends on the user interface, where the \reviewer accesses \applicant information by, for example, navigating through directories or downloading \applicant files. 
A system for admissions, for example, may be designed such that more clicks are needed to access different \applicants than different attributes within the same \applicant.
In this case, \holistic allocation may incur lower switching costs than \segmented allocation. 
However, in addition to the operational cost incurred by the user interface, another consideration relates to the cognitive load of switching between different types of information. 
For example, in assessing applicants for admissions, if an \reviewer operating in \holistic allocation has to shift from reviewing transcripts and test scores to evaluating essays and recommendation letters, the time and cognitive effort involved in this transition between attributes may outweigh the savings gained from the user interface. Consequently, whether \holistic or \segmented allocation leads to greater efficiency as a result of reduced switching costs depends on the user interface and the similarity in reasoning about different attributes.

\begin{hypothesis}
    (a) \Holistic allocation results in more efficiency than \segmented allocation, when transitioning from one \applicant to another requires more time or clicks than transitioning between attributes of the same \applicant.\\     
    (b) \Segmented allocation results in more efficiency than \holistic allocation, when transitioning from one attribute to another requires high cognitive effort due to the level of variation in the data and assessment process, taking more time than transitioning between \applicants for the same attribute.
\end{hypothesis}

We remark that the user interface should be designed to support the chosen allocation scheme. Specifically, if \segmented allocation is used, then the interface should be constructed so that the switching cost between \applicants for the same attribute should be made as low as possible.

\subsection{Mitigating Bias}
A major concern that regularly arises in evaluation and selection processes is that of bias. Researchers consider a decision to be biased when there is deviation from what is normatively predicted by classical probability and utility theory to be the optimal outcome based on the information or options available~\cite{hilbert2012toward}. Bias in decision making is a widely-studied topic in a number of fields, as it has substantial implications not only for evaluation and selection decisions, but also for many other high-stakes applications such as medical diagnosis, crime prevention, and financial performance, to name just a few~\citep{saposnik2016cognitive,costa2017bibliometric,kovera2019racial}.

One type of biases of particular concern for evaluation and selection is those that result in systematic discrimination against certain groups on the basis of information that is irrelevant or inappropriate for assessment~\cite[Section 7]{bertrand2004call,moss2012science,tomkins2017reviewer,shah2022surveyextended}. Many biases operate on a subconscious level~\cite{greenwald2003understanding} and thus affect evaluations even when the \reviewer intends to be fair. Consequently, the common recommendations include limiting subjective human judgment by using objective measures wherever possible, or when humans are making subjective assessments, ensuring that those are guided by specific outcome-relevant criteria and structured for consistent application to each \applicant~\citep{campion1988structured,pogrebtsova2020selection}. 

We propose that the allocation scheme also has an impact on mitigating bias. Specifically, we anticipate that \holistic and \segmented allocations affect outcomes by limiting the impact of highly biased evaluators on overall decision accuracy, and restricting access to biasing information.

\subsubsection{Reducing the impact of biased evaluators.}
It is likely that different \reviewers are biased to different extents. When some \reviewers are biased and some are not (or less so), \holistic and \segmented allocations are likely to lead to different types of impact. In \holallo, any particular \applicant has a certain probability of being assigned a biased \reviewer (depending on the fraction of biased \reviewers). Consequently, a subset of the \applicants are be highly affected by biased decisions, while the rest of the \applicants are not. By contrast, in \segallo, the probability that all attributes of a particular \applicant are assessed by highly biased evaluators becomes lower; however, it is more likely that each \applicant receives some assessment from at least one biased evaluator, compared to \holallo.
    
\begin{hypothesis}[Studied in Section~\ref{sec:bias}]\label{h:fairness}
   Compared to \holallo, \segallo better mitigates the impact of biased \reviewers on the accuracy of the \applicant evaluation, by reducing the chances that all attributes of any particular \applicant are evaluated by biased \reviewers.
   \label{hypothesis-bias2}
\end{hypothesis}

\subsubsection{Restricting access to biasing information.} Arguably, many interventions that have been made over the last several decades in traditional evaluation and selection processes are focused on limiting evaluators' access to biasing information. One famous example comes from symphony orchestras as they made efforts to incorporate more female musicians in the 1970s and 1980s~\cite{goldin2000orchestrating}. Initial diversity efforts yielded limited progress, even when many orchestras conducted auditions using screens to block evaluators' view of the candidates. However, one observant evaluator noted the difference in the sound on the wooden stage floor as the musicians entered for their audition, particularly the distinct sound of the high heels worn by the female musicians in contrast to the flat sounds made by most men's dress shoes. Consequently, a number of groups began using a carpeted walkway in addition to the screen, which resulted in a sudden increase in the number of women invited to join \cite{goldin2000orchestrating}. In generalizing this idea to the context of evaluation and selection, we hypothesize that \segallo mitigates bias by limiting access to information about irrelevant and potentially biasing attributes. For example, an evaluator could be asked to evaluate the research statements of graduate school applicants without access to any other information about the applicants, substantially limiting the possibility of bias.
\begin{hypothesis}\label{hypothesis:bias_access}
    \Segmented allocation helps mitigate the impact of bias compared to \holistic allocation, as a result of limiting evaluators' access to biasing information when they assess individual attributes of the \applicants.
\end{hypothesis}

\section{Modeling Framework}
We describe the mathematical framework used in our analysis.

\paragraph{Notation.} We assume that there are $\numapps$ \applicants, and each \applicant has $\numattrs$ attributes. We let $\cellmtx_{\idxapp\idxattr}\in \reals$ be the true quality of \applicant $\idxapp\in [\numapps]$ on attribute $\idxattr\in [\numattrs]$.\footnote{
    We use the notation $[\kappa]\defn \{1, 2, \ldots, \kappa\}$ for any positive integer $\kappa$.
} A higher value represents higher quality.
When there is more than one attribute, we define the true ranking of the applicants as the ranking induced by the mean of their attribute values.
The evaluation task is represented by the matrix $\{\cellmtx_\idxpair\}_{\idxapp\in [\numapps], \idxattr\in [\numattrs]}$, and we divide the matrix into sub-matrices as shown in Figure~\ref{fig:scheme}, where each \reviewer assesses a smaller sub-matrix consisting of a subset of the \applicants and a subset of the attributes (where the subset is allowed to equal the entire set). For simplicity, we assume each attribute of each \applicant is evaluated once, so all the sub-matrices are disjoint and collectively partition the entire matrix. We let $\score_{\idxapp\idxattr}\in \reals$ denote the score given to attribute $\idxattr$ of \applicant $\idxapp$ by the assigned \reviewer. Note that $\score_{\idxapp\idxattr}$ is often a noisy evaluation of $\cellmtx_{\idxapp\idxattr}$.

\paragraph{Metric.} In many evaluation and selection processes such as hiring or academic admissions, the goal is to choose a specified number of \applicants of the highest quality. Therefore, the accuracy of the evaluation process is determined by the top-K accuracy in ranking. For simplicity, we consider the top-$1$ accuracy as studied by~\citet{kleinberg2018rooney}. That is, the accuracy is $1$ if the estimated ranking correctly identifies the best \applicant in the true ranking, and $0$ otherwise.\footnote{
    In our setup, we make sure that there exists a unique best applicant in the true ranking. If there are ties in the estimated ranking, the accuracy is computed as $1 / \text{(number of \applicants in the tie)}$ if the true best \applicant is one of the estimated \applicants in the tie, and $0$ otherwise.
} We also consider a second error of metric that is suitable for understanding the calibration of \reviewers. This error represents the mean error in estimating the percentile of each \applicant, described in detail in Section~\ref{sec:calibration}.

\paragraph{Data generation.} In our simulations, we follow prior work~\cite{kleinberg2018rooney} and generate the attribute values from the power-law distribution unless specified otherwise. The power-law distribution with parameter $\parampowerlaw > 0$ is defined as $\Prob[Z \ge t] = t^{-(1+\parampowerlaw)}$ supported on $t\in [1, \infty)$, where $Z$ denotes the random variable.

We allow the attributes to be correlated, defined by a correlation parameter $\corr\in [-1, 1]$. For any desired distribution with c.d.f. $F$, we define the following procedure (cf.~\citealp{nelsen2010copula}) to generate $\numattrs$-dimensional correlated random variables. Let $\Phi$ denote the c.d.f. of the standard normal. For each applicant $\idxapp$, we first sample a vector $\vecpercentiles_\idxapp\in \reals^\numattrs$ from a multinomial normal distribution as $\vecpercentiles_\idxapp\sample \normal(0, (1-\corr)I_\numattrs + \corr\vecone_\numattrs\vecone_\numattrs^T)$ independent across the applicants $\idxapp\in [\numapps]$. Then we compute the attribute values as
$\cellmtx_{\idxapp\idxattr} = F^{-1}(\Phi(z_{\idxapp\idxattr}))$. 
It can be verified that each $\cellmtx_{\idxapp\idxattr}$ has marginal distribution $F$.  As special cases, when $\corr=1$, all attributes have identical values; when $\corr=0$, all attributes are independent.

\section{Methods and Results}
In this section, we examine our hypotheses related to calibration, efficiency and mitigating bias. 

\subsection{Calibration}\label{sec:calibration}
We focus on studying the relation between calibration accuracy and the number of \applicants assigned to an \reviewer, as described by Hypothesis~\ref{hypothesis-accuracy}.

\paragraph{Operationalization of calibration.}
Formally, we define calibration as the \reviewer's accuracy of estimating the ranking (or percentile) of each \applicant with respect to the entire pool of all \applicants. We define calibration on this relative scale for three reasons. First, the selection problem is intrinsically relative in nature, that is, we aim to select the top \applicants compared to the entire pool. Second, in many applications, the \reviewers are asked to report relative data. For example, \reviewers may be asked to give scores on a scale of $1\hyphen5$, where the criteria define the score of $1$ as the \applicant being the bottom 20\% among all \applicants, and $2$ as being 20\hyphen40\% among all \applicants, etc. Third, social comparison theory suggests that people's reasoning has a relative nature~\cite{festinger1954social}. For example, being a ``top'' \applicant is perceived as simply being significantly better than the rest of the \applicants. For this reason, using a relative scale than an absolute scale is shown to be more effective in various judgment tasks~\cite{goffin2011relative}. 

\paragraph{Experimental setup.}

To isolate the impact of calibration, we make a number of design simplifications, and conduct an experiment focusing on a single \attribute. 
We recruit $200$ crowdsourcing workers on the Prolific platform. The workers are introduced to a hiring context and asked to evaluate scores of \applicants. Specifically, they are told that there are $1000$ \applicants with scores that are integers between $0$ and $300$, without any distributional information about the scores. Then the workers are presented some numbers in between $200$ and $300$, and are asked to estimate the percentile of the scores. The workers classify each score to one of the five bins with respect to the population: $0\hyphen20\%$, ${20}\hyphen{40}\%$, ${40}\hyphen{60}\%$, ${60}\hyphen{80}\%$, and $80\hyphen100\%$. We choose to ask the workers to report in $5$ quantized bins instead of directly reporting a number of percentile, because prior studies have shown that workers are not able to perceive fine numbers accurately due to limited processing abilities~\cite{miller1956magic,shah2016estimation} and therefore have higher accuracy when a small number of quantized choices are given (e.g.,~\citealp{lietz2010questionnaire}). We have confirmed this trend by a preliminary study comparing using $5$ bins versus $10$ bins. 

\paragraph{Question grouping.} The workers are divided into two groups uniformly at random. Recall that there is a single attribute. In the first group, each worker is presented scores of $5$ \applicants (termed ``\groupfive''). 
In the second group, each worker is presented scores of $20$ \applicants (termed ``\grouptwenty''). The workers are always presented with $5$ scores per page. That is, for the \grouptwenty, the $20$ questions are distributed across $4$ pages. Neither group of workers is told the number of scores they will be presented before starting the task. 
The workers are required to answer all questions on a page before proceeding to the next page, though they are allowed to review and edit their answers on previous pages at any time before submission.
We choose to present $5$ questions per page and not inform the workers the total number of questions, to address the confounder that a worker who knows they have to do 20 questions may put less effort per question than if they knew  they have to do only 5 questions.

\paragraph{Values of scores.}
Since we consider a single attribute, we use the shorthand $\cellmtx_\idxapp \defn \cellmtx_{\idxapp 1}$ for the true score of each \applicant $\idxapp$. Let $F$ be the distribution $\normal(230, 25)$, truncated to the range of $[200, 300]$. The scores $\{\cellmtx_{\idxapp}\}_{\idxapp\in [\numapps]}$ in the \grouptwenty are generated i.i.d. from $F$. We pair up workers in the \grouptwenty and the \groupfive. For the scores in the \groupfive, we use the same values as the last $5$ questions in the \grouptwenty for a direct comparison. We choose this distribution for scores, because in a preliminary study where the workers are presented scores in the range of $[0, 100]$, we observe that the workers appear to have a strong uniform prior, mapping scores in $[0, 20]$ to percentile $0\hyphen20\%$, etc. This uniform mapping is an artifact of the experimental design that the quality under evaluation is real-valued. In more realistic situations, such a simplified mapping, say from \applicants' interview performance  to scores, does not exist. We therefore choose a range that is not $[0, 100]$ so that the workers do not rely on such priors.

\paragraph{Experimental Results.}
We record the worker calibration measured by their accuracy in estimating the percentile bins. Formally, let $\funcbin$ be the function mapping the percentile $0\hyphen 20\%, 20\hyphen40\%, 40\hyphen60\%, 60\hyphen80\%$ and $80\hyphen100\%$ to the bins $1, 2, 3, 4$ and $5$, respectively. For a single worker, let $y_\idxapp\in [5]$ be the bin reported for \applicant $\idxapp$. Then the absolute error between the true bin and the reported bin for \applicant $\idxapp$ incurred by this worker is defined as $\abs*{\funcbin(F^{-1}(\cellmtx_\idxapp)) - y_\idxapp}$.

For each worker, we compute their mean error over the \applicants they evaluate.
The workers' mean error is $1.14 \pm 0.06$ in the 5Q-group, and $0.84\pm 0.05$ in the 20Q-group. 
We perform a univariate permutation test between the mean errors of workers in the \grouptwenty, and those of workers in the \groupfive, using the difference in sample means as the test statistic.
We reject the null hypothesis that the errors from the two groups have the same mean (one-sided $p$-value $<0.01$; Cohen's effect size $d=0.52$). This result indicates that evaluation in the \grouptwenty is more accurate than in the \groupfive, confirming Hypothesis~\ref{hypothesis-accuracy} that \reviewers have better calibration when they see more \applicants.

\begin{figure}[tb]
    \centering
    \includegraphics[width=\figwidth]{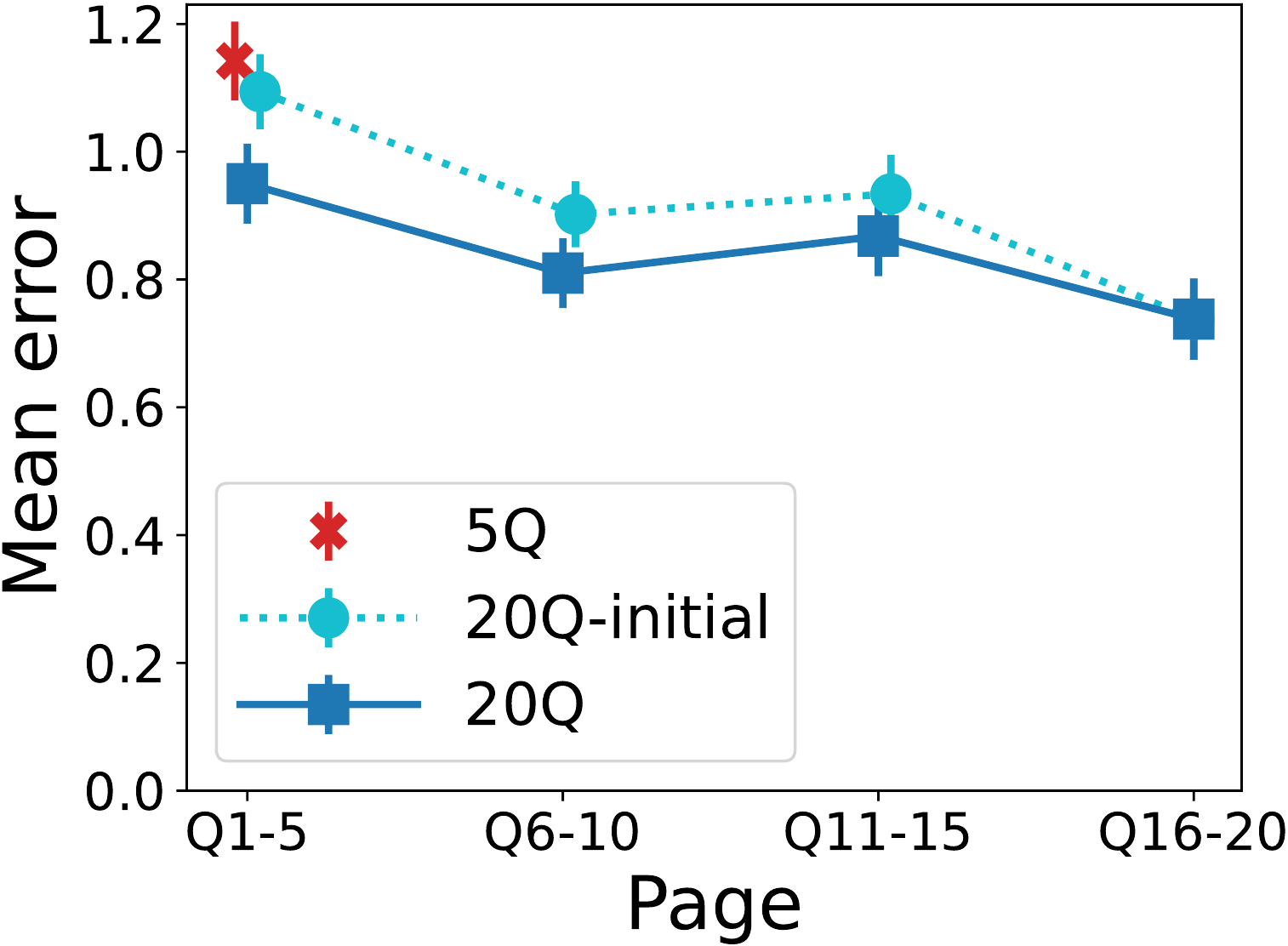}

    \caption{The mean error in estimating the percentile bins, for workers in the \groupfive (representing holistic allocation) and the \grouptwenty (representing segmented allocation). Error bars represent the standard error of the mean.
    }
    \label{fig:calibration}
\end{figure}

For the \grouptwenty, we also separately compute each worker's mean error over each page of $5$ questions (that is, Q1-5, Q6-10, Q11-15, Q16-20). The mean error for each page is plotted in Figure~\ref{fig:calibration}. For the \grouptwenty, we also plot the error on each page using the answers reported right before the workers ever turn to see the next page (see the curve ``\grouptwentyinit'').
The difference between the curves of the initial and final errors thus corresponds to the gain in calibration by workers correcting their answers to previous \applicants by seeing \applicants from later pages.
First, we observe that such corrections notably decrease the error, especially for the first page. This observation further provides evidence for Hypothesis~\ref{hypothesis-accuracy} by showing that workers are able to use the information they see from \applicants to perform correction.
Second, even after this correction, the error has a decreasing trend from earlier pages to later pages, suggesting that workers have limited abilities in performing such corrections. Specifically, in the \grouptwenty, the (final) mean error for page $1$ is $0.95 \pm 0.06$, and the (final) mean error for page $4$ is $0.74 \pm 0.06$. We perform a univariate permutation test between the mean errors for page $1$ and page $4$, using the difference in sample mean as the test statistics. We reject the null hypothesis that the errors for these two pages have the same mean (one-sided $p$-value $<0.01$; Cohen's effect size $d=0.34$).
Third, as a sanity check, we observe that for page $1$, the mean error in the \grouptwentyinit curve is similar to the mean error of the \groupfive. This is expected, as the workers from the two groups have strictly the same information before the workers in the \grouptwenty ever turn to the second page. 

We  observe the same qualitative trends in a previous version of the experiment, discussed in Appendix~\ref{app:previous_expt}.

\subsubsection{Simulations.}
The key observation from the crowdsourcing experiment is that seeing more \applicants improves calibration. We now conduct additional simulations for a more quantitative understanding. 
We stick with the setting of a single attribute.
We consider the following model for \reviewers. When an \reviewer is assigned $\numapps$ \applicants, it assigns the lowest $\frac{\numapps}{5}$ \applicants to the bin $0\hyphen20\%$, followed by the next $\frac{\numapps}{5}$ \applicants to the bin $20\hyphen40\%$, etc. This is a natural model for \reviewers, because as $\numapps$ goes to infinity, the mean error on the reported bins approaches $0$.

\begin{figure}[tb]
    \centering
    \includegraphics[width=\figwidth]{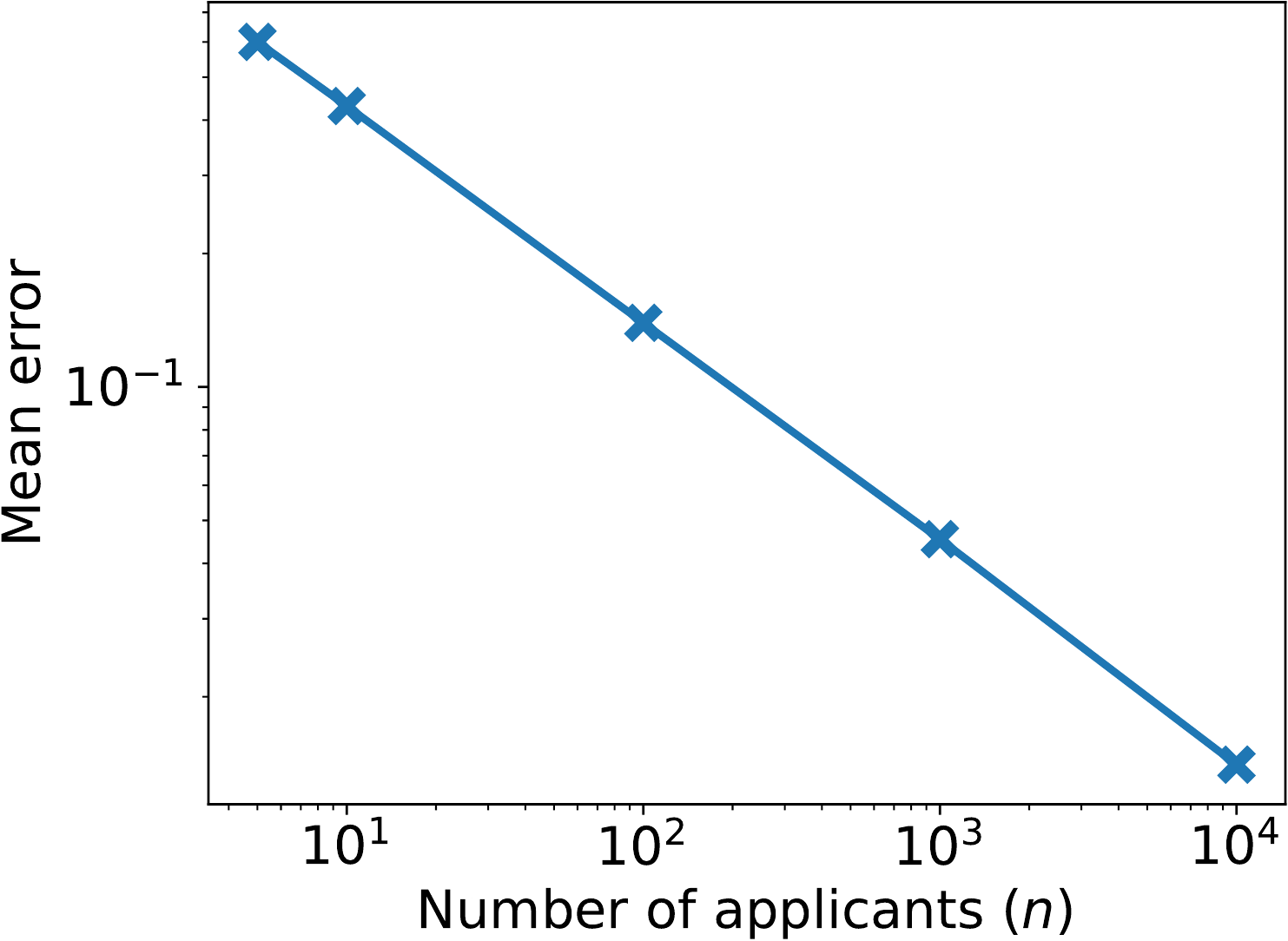}\label{float:calibration_l1_vs_n}
    \caption{The mean error in calibration of a single \reviewer, as a function of the number of \applicants. Each point is computed over $1000$ runs (error bars are too small to be visible).}
    \label{fig:calibration_simulation}
\end{figure}

We plot the mean error as a function of the number of \applicants $\numapps$ assigned to a single \reviewer in Figure~\ref{fig:calibration_simulation}.
The error decreases as the number of applications $\numapps$ increases, matching the experimental result and therefore providing additional evidence supporting Hypothesis~\ref{hypothesis-accuracy}. We empirically observe that as the number of \applicants $\numapps$ increases, the mean error decreases at a rate of  $\frac{1}{\sqrt{\numapps}}$.

\subsection{Efficiency}\label{sec:efficiency}
We study the adaptive allocation of effort in Hypothesis~\ref{hypothesis:efficiency} via simulations.

\paragraph{Setting.} We consider $\numapps=200$ \applicants, and for simplicity, $\numattrs=2$ attributes assessed by two \reviewers. In segmented allocation, each \reviewer is assigned one attribute of all \applicants. In \holistic allocation, each \reviewer is assigned both attributes of half of the \applicants. 
The attribute values are generated from a power-law distribution with parameter $1$, with correlation $\corr\in [0, 1]$ between the two attributes. To isolate the efficiency aspect from calibration errors, we assume that an \reviewer always reports the true value of the attributes, namely $y_{\idxapp\idxattr} = \cellmtx_{\idxapp\idxattr}$ for each $(\idxapp, \idxattr)$ pair.

According to Hypothesis~\ref{hypothesis:efficiency}, \holistic allocation provides the opportunity for an \reviewer to decide whether to evaluate the second attribute of an applicant, based on the quality of the first attribute. For simplicity, we assume that in \holistic allocation, each \reviewer always reviews attribute $1$ of all \applicants. 
Each \reviewer then reviews attribute $2$ only on the \applicants who have scored high on attribute $1$. Specifically, we assume that attribute $2$ is only evaluated on a $\thresh$-fraction\footnote{
    Selecting the top $\thresh$-fraction requires knowledge about attribute $1$ of all the applicants that an \reviewer is assigned. In practice, an \reviewer may select the applicants whose attribute $1$  exceeds a certain real-valued threshold, which approximately has the same effect.
} of the \applicants receiving the top scores on attribute $1$, for a parameter $\thresh\in (0, 1]$. Finally, the best applicant is selected as the one whose mean of the two attribute scores is the maximum, namely $\argmax_{\idxapp\in [\numapps]} (\score_{\idxapp 1} + \score_{\idxapp 2})$, from the applicants on which both attributes are evaluated.


\begin{figure}[tb]
    \centering
    \includegraphics[width=\figwidth]{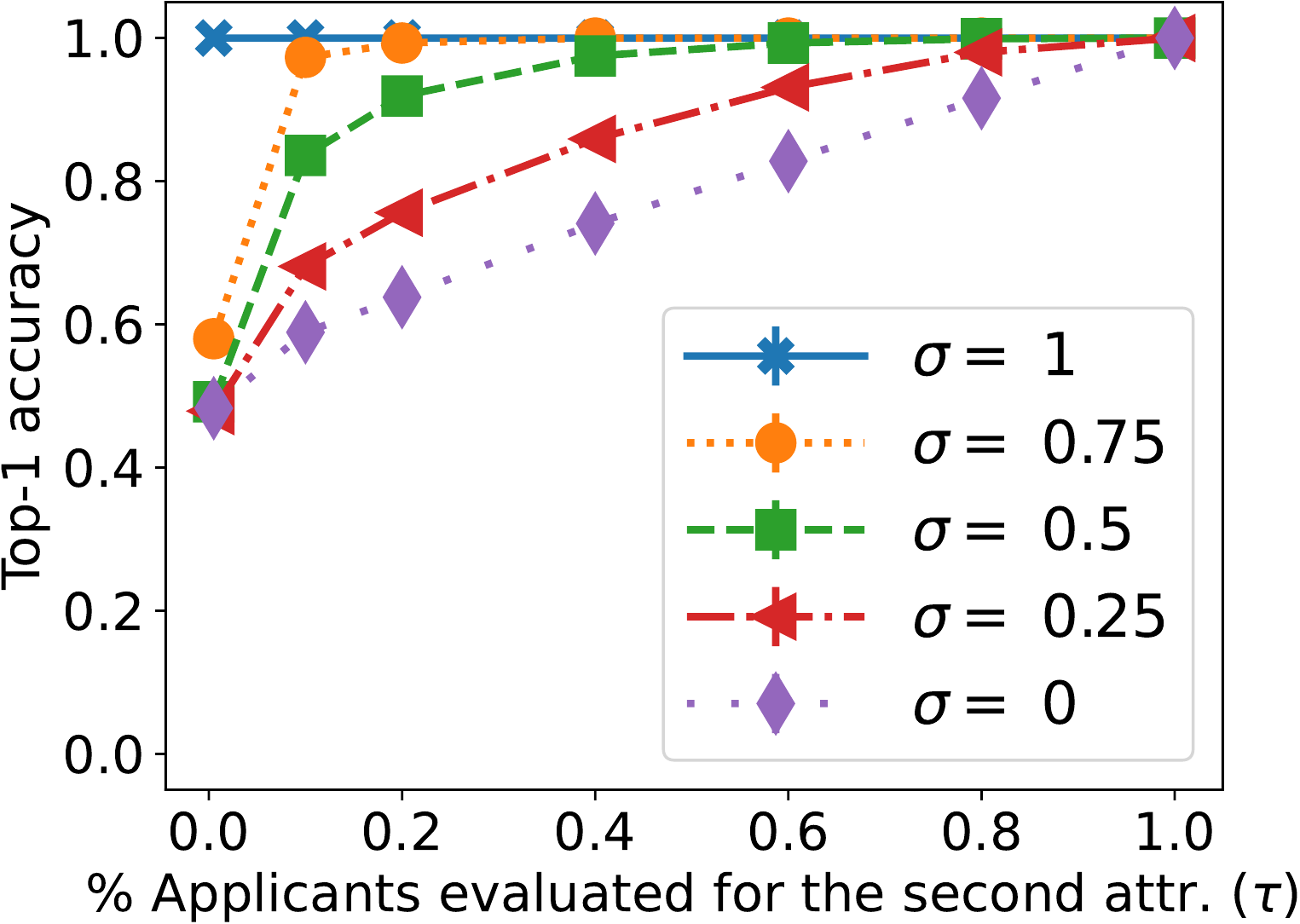}
    \caption{
        Top-$1$ accuracy for different fractions $\thresh$ of the \applicants evaluated for the second attribute, and various values of the correlation $\corr$ between the two attributes.
        Each point is computed over $1000$ runs (error bars are too small to be visible).
    }
    \label{fig:cutoff}
\end{figure}

\paragraph{Simulations.} 
In Figure~\ref{fig:cutoff}, we compute the top-1 accuracy for different fraction $\thresh$ and attribute correlation $\corr$.
When the correlation is $\corr=1$ (see the blue curve), by definition evaluating only attribute $1$ achieves perfect accuracy, and there is no need to evaluate attribute $2$. When the correlation $\corr$ is relatively high, we observe that relatively small values of $\thresh$ introduce a significant amount of saving in terms of the total number of attributes evaluated, while the accuracy only decreases marginally. This observation validates Hypothesis~\ref{hypothesis:efficiency}, and we conclude that a higher correlation between the attributes allows more saving in \holistic allocation. This result points to a tradeoff between efficiency and accuracy in \holistic allocation -- namely, smaller $\thresh$ introduces savings but also more error. The specific point to pick in this tradeoff depends on the goals of the system designer.

\subsection{Mitigating Bias}\label{sec:bias}
To study Hypothesis~\ref{hypothesis-bias2}, we present a simple model for analyzing the effect of bias reduction. We present theoretical guarantees and simulational results that characterize the regimes under which \segmented allocation results in more accurate and less biased evaluations than \holistic allocation. Our results provide intuition on the effect of redistributing and reducing the impact of biased \reviewers.

\paragraph{Formulation.}
Recall that $\cellmtx_{\idxapp\idxattr}$ denotes the true value of \applicant $\idxapp\in [\numapps]$ on attribute $\idxattr\in [\numattrs]$.
We assume that the \applicants consist of two groups -- advantaged and disadvantaged -- where a fraction $\fracappsdisadvantage\in [0, 1]$ of the \applicants are from the disadvantaged group. We assume that a fraction $\fracfeatsdisadvantage\in [0, 1]$ of the attributes are ``protected''. Each \reviewer has an independent probability of $\fracbiased\in (0, 1)$ to be biased in the following sense: An unbiased \reviewer reports the (noiseless) true value $\score_{\idxapp\idxattr} = \cellmtx_{\idxapp\idxattr}$ for any applicant $\idxapp$ and any attribute $\idxattr$ that they are assigned, while a biased \reviewer applies a multiplicative bias factor $\discount \in [0, 1)$ to the protected attributes of the disadvantaged applicants, and reports the true value otherwise. In other words, for attribute $\idxattr$ of applicant $\idxapp$, a biased \reviewer reports
\begin{align*}
    \score_{\idxapp\idxattr}=\begin{cases}
    \beta \cellmtx_\idxpair & \text{ if $\idxattr$ protected and $\idxapp$ disadvantaged}\\
\cellmtx_\idxpair & \text{otherwise}.
\end{cases}
\end{align*}
For ease of analysis, we consider a simple case of $\numattrs=2$ attributes, where the correlation between the two attributes is $\corr=1$. That is, for each applicant $\idxapp$, the two attributes have identical values $\cellmtx_{\idxapp1} = \cellmtx_{\idxapp2}$. We hence use the shorthand $\cellmtx_{\idxapp}$ to denote this value. 
We assume that the values $\{\cellmtx_\idxapp\}_{\idxapp\in [\numapps]}$ are generated i.i.d. from a continuous\footnote{
    We consider continuous distributions for simplicity, so that the best \applicant is uniquely defined with probability $1$.
} distribution $\distval$ supported on $[0, \infty)$, such as the power-law distribution.
Let $\setdisadvantage\subseteq [\numapps]$ denotes the set of $\fracappsdisadvantage\numapps$ disadvantaged \applicants, and let $\setadvantage\subseteq [\numapps]$ denotes the set of $(1-\fracappsdisadvantage)\numapps$ advantaged \applicants. Denote the quality of the best \applicant in the disadvantaged group by $\valmaxdisadvantage \defn \max_{\idxapp\in \setdisadvantage} \valdisadvantage_\idxapp$, and likewise denote $\valmaxadvantage\defn \max_{\idxapp\in \setadvantage} \valdisadvantage_\idxapp$. We compute the mean of attribute scores for each applicant, and estimate the best applicant by selecting the one with the maximum mean score. Denote the expected top-1 error under \holistic and \segmented allocations by $\errhol$ and $\errseg$ respectively, formally defined by $\Prob(\argmax_{\idxapp\in [\numapps]} \cellmtx_\idxapp \ne \argmax_{\idxapp\in [\numapps]}  \score_{\idxapp 1} + \score_{\idxapp 2})$ , using the scores $\{\score_\idxpair\}$ under the two allocation schemes respectively. 

\subsubsection{Theoretical results.} We focus on a simplified case of two \reviewers, which as we see shortly, already illustrates the intricacy of the comparison. In this setting, \holallo assigns each \reviewer both attributes of half of the applicants; \segallo assigns each \reviewer one attribute of all applicants.
We assume that the assignment to \applicants and attributes is uniformly at random.

\begin{theorem}\label{prop:compare_seg_hol_two_attrs}
Let the number of attributes be $\numattrs=2$. Let the fraction of disadvantaged \applicants be $\fracappsdisadvantage= 0.5$. Let the two attributes have identical values (that is, $\cellmtx_\idxapp\defn \cellmtx_{\idxapp1} = \cellmtx_{\idxapp2}$), sampled i.i.d. from a continuous distribution $\distval$. Consider \holistic and \segmented allocations under two \reviewers.

\begin{enumerate}[label=(\alph*)]
    \item \label{item:attrs_one}
    Let $\fracfeatsdisadvantage = 0.5$, that is, one of the two attributes is protected. Then for any bias factor $\discount\in [0, 1)$ and any \reviewer bias probability  $\fracbiased\in (0, 1)$, segmented allocation incurs a lower error than \holistic allocation, that is, $ \errseg \le \errhol$.

    \item \label{item:attrs_both}
    Let $\fracfeatsdisadvantage = 1$, that is, both attributes are protected. Let $\discount=0$ (extreme downward bias against disadvantaged \applicants). Then
    \begin{align}
         \errhol - \errseg = \frac{\fracbiased(1-\fracbiased)}{2} \left[4\cdot\Probbig{\valmaxdisadvantage > 2\valmaxadvantage}-1\right].\label{eq:prop_err_comparison}
    \end{align}
    Hence, for any $\fracbiased \in (0, 1)$, \segallo incurs a lower error than \holallo, if and only if
    \begin{align}\label{eq:condition}
        \Probbig{\valmaxdisadvantage > 2 \valmaxadvantage} > 0.25.
    \end{align}
    This condition~\eqref{eq:condition} is dependent on the number of applicants $\numapps$ and and the distribution $\distval$, and independent of the other problem parameters. 
    In particular, for the power-law distribution with a constant parameter $\parampowerlaw$, \segallo is better than \holallo for sufficiently large $\numapps$, if and only if \begin{align}
        \parampowerlaw < \frac{\log(3)}{\log(2)} - 1\approx 0.58.\label{eq:condition_powerlaw}
    \end{align}
\end{enumerate}
\end{theorem}

The proof of this theorem is provided in Appendix~\ref{app:proof}. This theorem reveals that \segallo is better than \holallo in terms of accuracy over a large range of parameters, but not always.
Despite the simplified settings considered in the theorem, the result illustrates how allocating biased \reviewers differently leads to changes in accuracy.

\subsubsection{Simulations.}

\input{text_fig_bias}

We study the effect of the set of parameters $(\parampowerlaw, \corr, \discount,\fracappsdisadvantage,\fracfeatsdisadvantage)$ in the model. Following the assumption of~\Cref{prop:compare_seg_hol_two_attrs}, we consider two \reviewers for simulation. The proof of~\Cref{prop:compare_seg_hol_two_attrs} suggests that it suffices to consider one biased \reviewer and one unbiased \reviewer.
We fix the number of applicants $\numapps = 20$ and the number of attributes $\numattrs = 20$. To inspect the difference between \holistic and \segmented allocations, for ease of visualization, we vary two parameters at a time while keeping the other ones fixed. For consistency, one varying parameter is always $\parampowerlaw$ for the power-law distribution. We set the default parameter values as $\corr=0.5$, $\discount = 0$, $\fracappsdisadvantage=0.5$ and $\fracfeatsdisadvantage = 1$, when they are not chosen as the parameter to be varied. The results are shown in Figure~\ref{fig:fairness} and discussed below.

\paragraph{Effect of power-law parameter ($\parampowerlaw$)}
In Figure~\ref{fig:fairness}\subref{float:fairness_sigma}-\subref{float:fairness_lda}, we observe the general trend that both \segmented and \holistic allocations achieve higher accuracy under smaller values of $\parampowerlaw$. A smaller $\parampowerlaw$ means that the distribution has a heavier tail, so that the values of the \applicants are more spread out. Hence, the best \applicant has a more extremal, higher value compared to the other \applicants, giving stronger signal for the evaluation process and making it easier.
Two exceptions to this general trend are \holallo in Figure~\ref{fig:fairness}\subref{float:fairness_sigma} and~\ref{fig:fairness}\subref{float:fairness_alpha}, where the accuracy is independent of $\parampowerlaw$. In these two cases, we have  $\fracfeatsdisadvantage=1$ and $\discount=0$. Hence, when a biased \reviewer is assigned a disadvantaged applicant in \holistic allocation, all attributes ($\fracfeatsdisadvantage=1$) are discounted to zero ($\discount=0$), making it impossible for disadvantaged \applicants to be identified as the best regardless of their values, and thus the accuracy is independent of $\parampowerlaw$.

\paragraph{Effect of correlation ($\corr$): Figure~\ref{fig:fairness}\subref{float:fairness_sigma}}

In \holallo, for the same reason that the accuracy is independent of $\parampowerlaw$ as previously explained, the accuracy is also independent of $\corr$. In \segallo, we observe that a higher correlation leads to a higher accuracy. This is because a higher (positive) correlation strengthens the signal for \applicants. For example, consider the extreme case of $\corr=1$. Then the same attribute value is replicated $\numattrs$ times for each applicant, improving robustness against randomness in the evaluation process due to bias.

Comparing \segmented and \holistic allocations, we observe that \segmented allocation performs better when $\corr$ is high (more correlation) and $\parampowerlaw$ is small (heavy tail in the distribution). 
The tradeoff between the two allocation schemes arises, because \segmented allocation always discriminates disadvantaged \applicants but to a lesser extent, whereas \holistic allocation discriminates disadvantaged \applicants less often but to a greater extent. When the correlation $\corr$ between the attributes is high, the gain from only discriminating a fraction of the attributes (as supposed to all attributes) is more significant.

Finally, note that in~\Cref{prop:compare_seg_hol_two_attrs} we set the correlation as $\corr = 1$. Hence, the setting of~\Cref{prop:compare_seg_hol_two_attrs}\ref{item:attrs_both} corresponds to the top most matrix row in Figure~\ref{fig:fairness}\subref{float:fairness_sigma}. We observe that the sign of the comparison between the two schemes is consistent with the theoretical result, with a change-point at $\parampowerlaw\approx 0.6$ in the right panel of Figure~\ref{fig:fairness}\subref{float:fairness_sigma}.

\paragraph{Effect of bias factor ($\beta$): Figure~\ref{fig:fairness}\subref{float:fairness_beta}}

We observe that both allocation schemes have a higher accuracy when the value of $\discount$ is large. This is natural because a larger $\discount$ corresponds to less discrimination by the biased \reviewers. Comparing the two schemes, \segmented allocation is more advantageous when $\discount$ is larger. We reason that when $\discount$ is small (more discrimination), the effect when a disadvantaged \applicant is discriminated is very detrimental for the \applicant (either on one attribute for \segmented allocation, or both attributes for \holistic allocation). Hence, \holistic allocation performs better because the probability that a disadvantaged \applicant is discriminated (to any extent) is smaller. On the other hand, when $\discount$ is large (less discrimination), the advantage of \segmented allocation of discounting only on one attribute becomes more significant, as supposed to discounting both attributes in \holistic allocation. 

\paragraph{Effect of fraction of disadvantaged \applicants ($\fracappsdisadvantage$): Figure~\ref{fig:fairness}\subref{float:fairness_alpha}}

We observe that segmented allocation performs better in general when $\fracappsdisadvantage$ is large. To reason about this effect, let us first think about the extreme case when $\fracappsdisadvantage=1$.
In this case, segmented allocation is better because it gives consistent treatment to all applicants. Namely, the biased \reviewer discounts one attribute of all applicants. On the other hand, \holistic allocation creates discrepancy between \applicants, because only the disadvantaged \applicants assigned to the biased \reviewer are discounted.
Moreover, note that the performance of \segmented allocation is not monotonic in $\fracappsdisadvantage$: For larger values of $\parampowerlaw$, \segmented allocation has the lowest accuracy when a large fraction, but not all applicants are disadvantaged. This non-monotonicity of \segmented allocation leads to the non-monotonicity in comparing the two schemes in the right panel of Figure~\ref{fig:fairness}\subref{float:fairness_alpha}.

\paragraph{Effect of fraction of protected attributes ($\fracfeatsdisadvantage$): Figure~\ref{fig:fairness}\subref{float:fairness_lda}}

We observe that segmented allocation performs better when  $\fracfeatsdisadvantage$ is small. In this case, there is less discrimination in both allocations: \Segmented allocation decreases the probability that a biased \reviewer is assigned a protected attribute, whereas \holistic allocation decreases the impact of a biased \reviewer on an \applicant. Our empirical observation aligns with the theoretical results: Comparing part~\ref{item:attrs_one} and part~\ref{item:attrs_both} of~\Cref{prop:compare_seg_hol_two_attrs} also suggests that segmented allocation performs better for smaller values of $\fracfeatsdisadvantage$.

\medskip 

In summary, there is a tradeoff where more segmentation means that the disadvantaged applicants are more likely to be consistently discriminated, but to a lesser extent; the parameters tip this tradeoff in different manners. We conclude that Hypothesis~\ref{hypothesis-bias2} does not capture the complete picture, as the benefit of \segmented allocation depends on the specific values of the parameters.\footnote{
    While we aim to provide visualization for the combined effect of parameters $(\corr, \fracappsdisadvantage, \fracfeatsdisadvantage, \discount, \parampowerlaw)$, the 2-dimensional cross sections in Figure~\ref{fig:fairness} are not meant to provide the complete picture, and the trends we observe should not be interpreted as holding universally when the fixed parameters are set to any arbitrary values. For example, when one parameter lies in a regime that strongly favors one allocation over the other, it is possible that changing other parameters do not change the sign of comparison in contrast to the sign changes shown in Figure~\ref{fig:fairness}.
}

\section{Discussion}\label{sec:conclusion}

In this work, we consider using segmented allocation as an alternative to the conventional \holistic allocation, for applications such as hiring and admissions. We provide detailed discussions comparing the two allocation schemes, and present theoretical and experimental results focused on three aspects: calibration, efficiency and fairness. These results indicate the potential improvement by \segallo on calibration, while also suggesting that \holallo has potential benefits on efficiency. The two allocation schemes also distribute \reviewers differently that lead to different impacts in terms of fairness. These results together suggest a tradeoff between \holistic and \segmented allocations (and the spectrum in between). The tradeoff and the choice of which allocation to use depends on the characteristics of specific applications and which dimensions are prioritized by the system designer. 

Immediate open problems include validating the remaining three hypotheses that are not analyzed in this paper, and extending the theoretical and simulation results to more general scenarios to improve our understanding of the bias considerations in Section~\ref{sec:bias}.
For example, if each attribute of each application is evaluated by many \reviewers, then it is natural to expect that the bias is averaged out more evenly across \reviewers, and the discrepancy between \holistic and \segmented allocations becomes less prominent.
There are also various other considerations, as well as open problems:

\begin{itemize}
    \item \Segallo requires grouping of attributes, and the system designer needs to do this grouping appropriately. 
    For example, in the case of admissions, one may group test scores and GPAs as one \attribute called ``scholarly performance''. In order to provide appropriate context to \reviewers, one may also need to provide the same attributes to multiple \reviewers. 
    
    \item In addition to grouping the attributes, it is also possible to group the \applicants. We have assumed that the \applicants are distributed to \reviewers uniformly at random. In reality, \reviewers may have different expertise that make them more suitable to review a particular subset of the \applicants. For example, in admissions, \reviewers from the same educational background as the \applicants may be more familiar with interpreting the schools and the GPAs. 
    
    \item We have assumed for simplicity that the final score is computed by taking the mean over all attribute scores. In practice, we may want to use different weights for different attributes, or even use non-linear functions. In some cases, the aggregation function may not be precisely provided by the system designer, but needs to be learned from past data. This problem of learning the aggregation function for evaluation has been studied in the specific context of peer review~\cite{noothigattu2018choosing}, and it is of interest to extend such results to more general applications.
    
    \item This work discusses a spectrum of choices in terms of the number of attributes and \applicants assigned to each \reviewer. An open problem of interest is to establish the optimal point(s) on this \holistic-\segmented spectrum theoretically and practically for any given specification of the applications and desiderata. 
\end{itemize}

\noindent\textbf{Acknowledgments.} We thank Lily Laredo for inputs on writing the worker instructions for the crowdsourcing experiments. This work was supported by the CMU Block Center and NSF CAREER 1942124.

%% file: text_fig_bias.tex
\newcommand{\figvspace}{3mm}
\begin{figure}[tb!]
\centering
    \hspace{6mm}
    \includegraphics[height=.45cm]{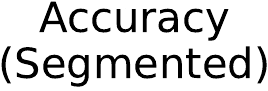}
    \hspace{10mm}
    \includegraphics[height=.45cm]{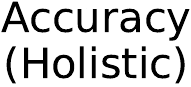}
    \hspace{10mm}
    \includegraphics[height=.45cm]{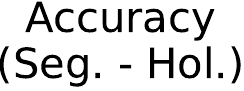}

    \hspace{12mm}
    \includegraphics[width=0.4\linewidth]{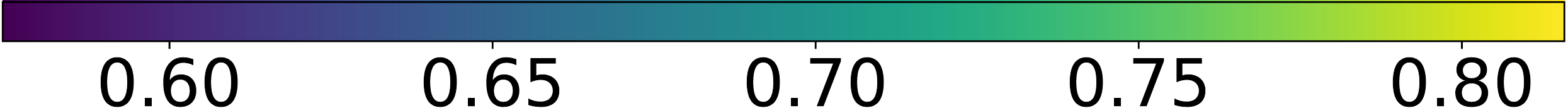}
    \hspace{7mm}
    \includegraphics[width=0.2\linewidth]{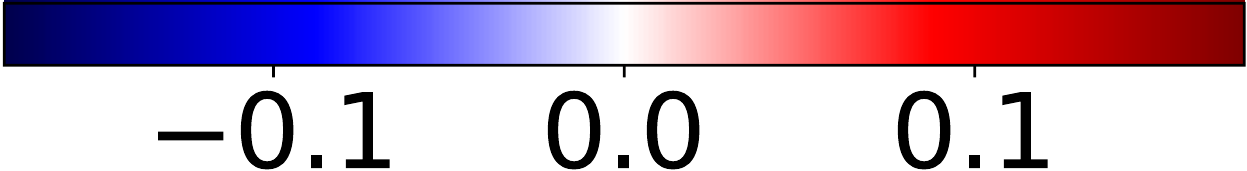}
    \subfloat[]{\label{float:fairness_sigma}
        \includegraphics[width=0.9\linewidth]{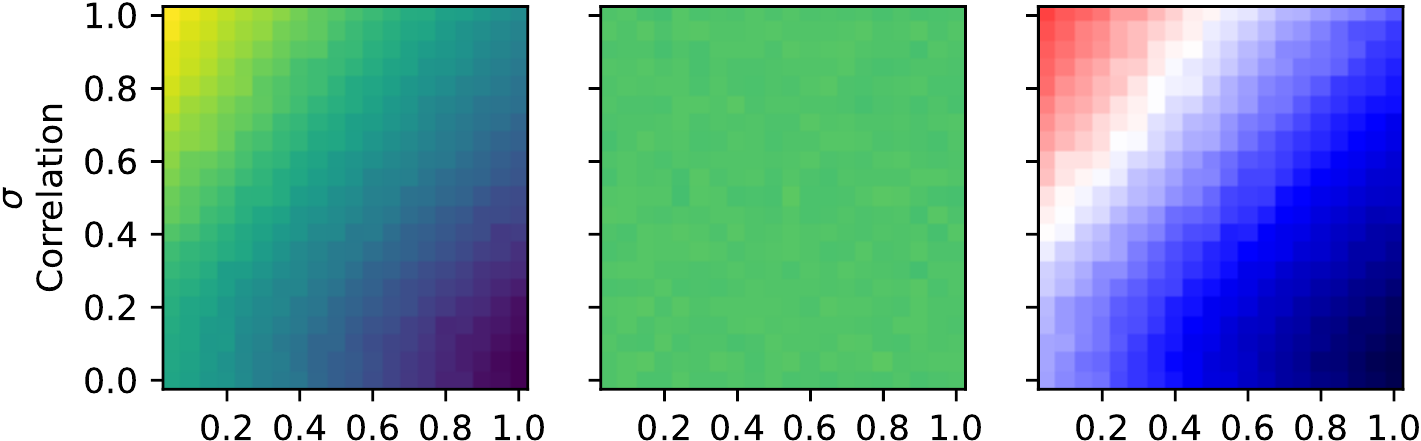}
        }
    \hrule\vspace{\figvspace}
    
    \hspace{12mm}
    \includegraphics[width=0.4\linewidth]{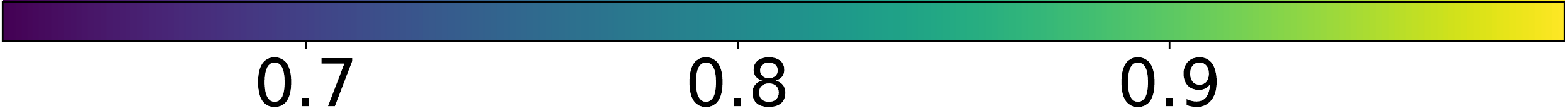}
    \hspace{7mm}
    \includegraphics[width=0.2\linewidth]{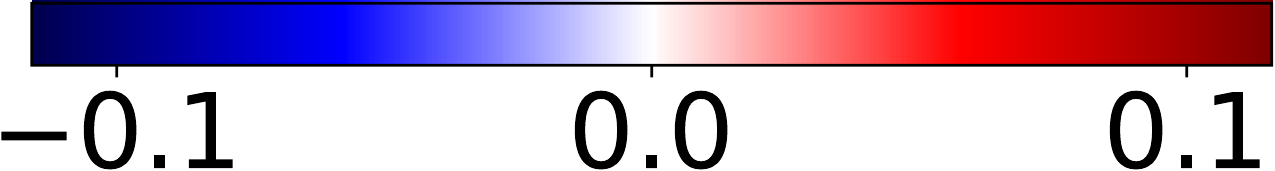}
    \subfloat[]{
        \includegraphics[width=0.9\linewidth]{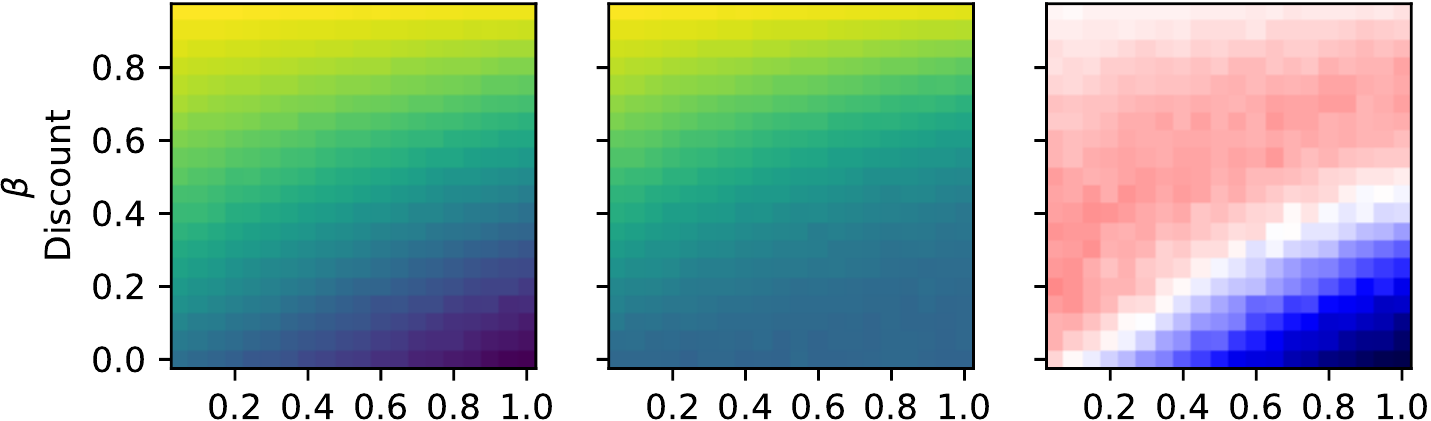}\label{float:fairness_beta}
    }
    \hrule
    \vspace{\figvspace}
    
    \hspace{12mm}
    \includegraphics[width=0.4\linewidth]{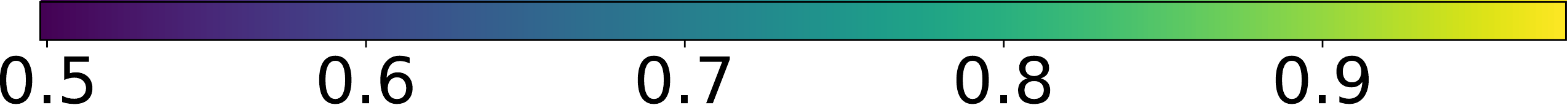}
    \hspace{7mm}
    \includegraphics[width=0.2\linewidth]{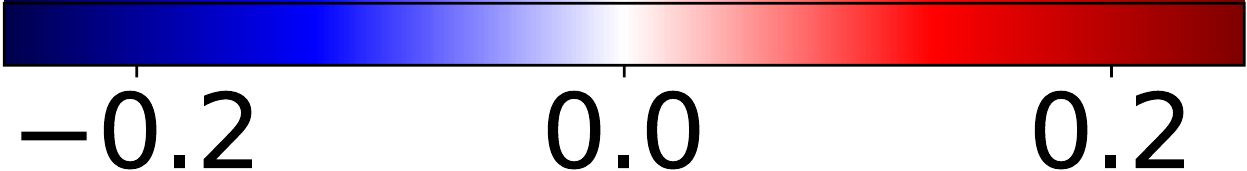}
    \subfloat[]{
        \includegraphics[width=0.9\linewidth]{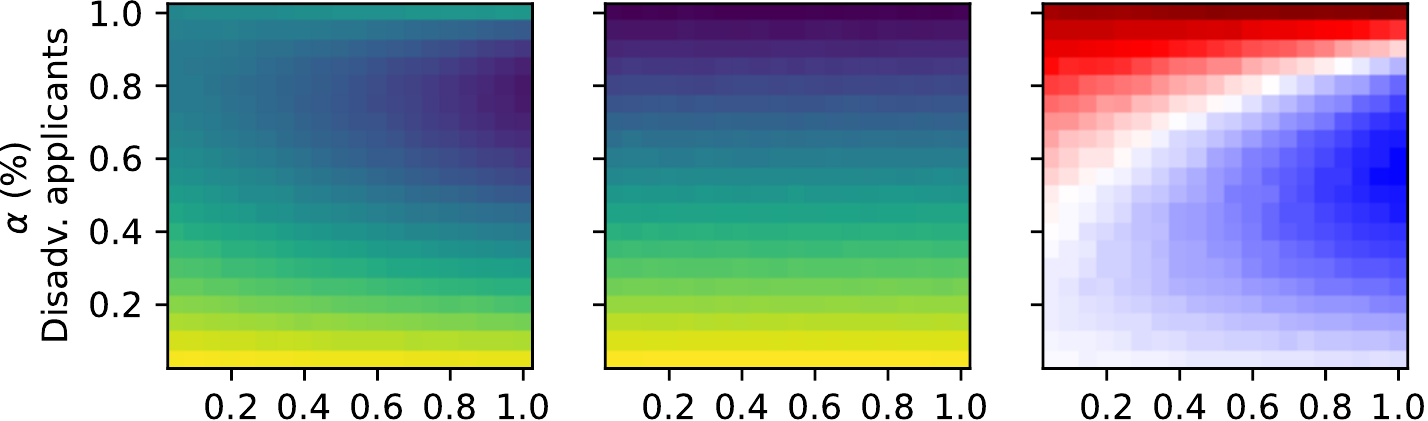}\label{float:fairness_alpha}
    }
    \hrule\vspace{\figvspace}
    
    \hspace{12mm}
    \includegraphics[width=0.4\linewidth]{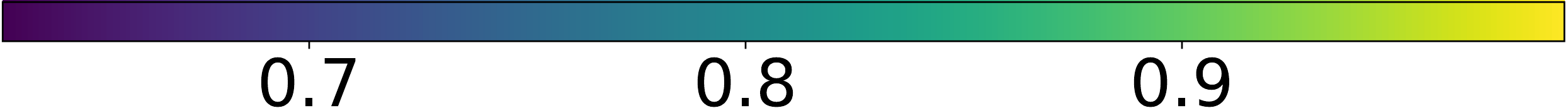}
    \hspace{7mm}
    \includegraphics[width=0.2\linewidth]{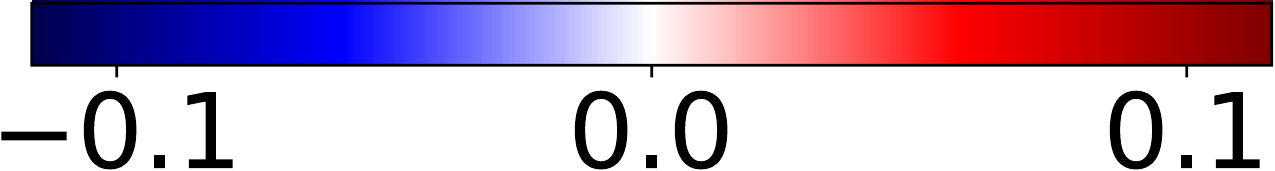}
    \subfloat[]{
        \includegraphics[width=0.9\linewidth]{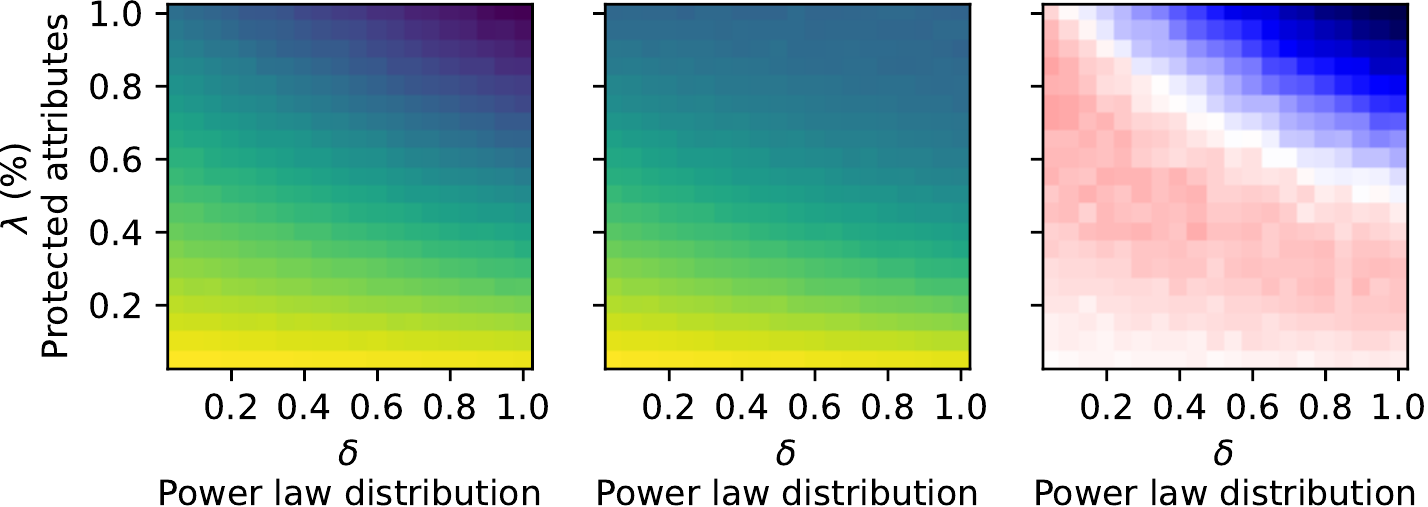}\label{float:fairness_lda}
    }

    \caption{[Best viewed in color.] Comparison between \holistic and \segmented allocations, under different parameter values for $\parampowerlaw$ and $(\corr, \discount, \fracappsdisadvantage, \fracfeatsdisadvantage)$. The plots depict the accuracy of segmented (left) and \holistic (middle) allocations, and their difference (right).
    For each choice of the parameters, the accuracy for the two allocation schemes is respectively computed over $50000$ runs.
    \label{fig:fairness}
}
    
\end{figure}

%% file: text_appendix.tex
\noindent{\bf \Large Appendices}

\section{Proof of Theorem~\ref{prop:compare_seg_hol_two_attrs}}\label{app:proof}
In this section, we present the proof of~\Cref{prop:compare_seg_hol_two_attrs}.
For any event $\event$, we let $\setcomplement{\event}$ denote the complement of $\event$. We first derive equality~\eqref{eq:proof_both_attrs_err_reduce_disadvantage} below that is common across parts~\ref{item:attrs_one} and~\ref{item:attrs_both}, and then separately prove for parts~\ref{item:attrs_one} and~\ref{item:attrs_both} based on equality~\eqref{eq:proof_both_attrs_err_reduce_disadvantage}.

Consider either part~\ref{item:attrs_one} or~\ref{item:attrs_both}, where one or both attributes are protected. Let $\eventerrorhol$ and $\eventerrorseg$ denote the events that the top-$1$ \applicant is identified incorrectly, for holistic allocation and segmented allocation, respectively. Formally, when the number of attributes is $\numattrs = 2$, the top-$1$ \applicant is identified incorrect when
\begin{align*}
    \argmax_{\idxapp\in [\numapps]} \cellmtx_{\idxapp} \ne \argmax_{\idxapp\in [\numapps]}\sum_{\idxattr\in [2]} (\score_{\idxapp 1} + \score_{\idxapp 2})
\end{align*}
under the respective allocation scheme. 
For each \applicant $\idxapp$, we term the mean of the attribute scores, $\frac{\score_{\idxapp 1} + \score_{\idxapp 2}}{2}$, as the estimated quality of the \applicant. We first observe that it suffices to consider the case where the best \applicant is disadvantaged: Since the bias factor $\discount$ only applies to the disadvantaged \applicants, for either holistic or segmented allocation, the estimated quality of the advantaged \applicants always equals their true quality; the estimated quality of the disadvantaged \applicants, due to the discounting, is always lower than or at most equal to their true quality. Hence, when the best \applicant is an advantaged \applicant, it is always identified correctly. Formally, recall that the random variables  $\valmaxadvantage$ and $\valmaxdisadvantage$ denote the true quality of the best \applicant in the advantaged and the disadvantage groups, respectively. For $\event\in \{\eventerrorhol, \eventerrorseg\}$, we have
\begin{align}
    \Prob(\eventerror\given \valmaxdisadvantage < \valmaxadvantage) = 0.\label{eq:proof_both_attrs_no_err_advantage_top}
\end{align}
Therefore, for $\event\in \{\eventerrorhol, \eventerrorseg\}$, we have
\begin{align}
    \Prob(\eventerror) & =
    \Prob(\eventerror \given \valmaxdisadvantage>\valmaxadvantage) \cdot \Prob(\valmaxdisadvantage>\valmaxadvantage) + \Prob(\eventerror \given \valmaxdisadvantage < \valmaxadvantage) \cdot \Prob(\valmaxdisadvantage< \valmaxadvantage)
    \nonumber\\
    & \stackrel{\stepone}{=} \Prob(\eventerror \given \valmaxdisadvantage>\valmaxadvantage) \cdot \Prob(\valmaxdisadvantage>\valmaxadvantage) \nonumber\\
    & \stackrel{\steptwo}{=} \frac{1}{2} \Prob(\eventerror \given \valmaxdisadvantage>\valmaxadvantage)\label{eq:proof_both_attrs_err_reduce_disadvantage},
\end{align}
where step~\stepone is true by equality~\eqref{eq:proof_both_attrs_no_err_advantage_top}; step~\steptwo is true because the fraction of disadvantaged \applicants is $\fracappsdisadvantage=0.5$, and hence $\Prob(\valmaxdisadvantage>\valmaxadvantage)=\frac{1}{2}$ by symmetry. Note that the true qualities $\valmaxdisadvantage$ and $\valmaxadvantage$ are generated from the continuous distribution $\distval$, so it is safe to ignore the case of $\valmaxadvantage=\valmaxdisadvantage$ which happens with probability $0$. 

We next observe that there is no difference between \segmented and \holistic allocations if neither or both reviewers are biased. Formally, let $\eventbiasedreviewer$ denote the event that exactly one out of the two reviewers is biased. We have
\begin{align*}
    \Prob(\eventerrorseg \given \valmaxdisadvantage > \valmaxadvantage, \setcomplement{\eventbiasedreviewer}) =  \Prob(\eventerrorhol \given \valmaxdisadvantage > \valmaxadvantage, \setcomplement{\eventbiasedreviewer}).
\end{align*}
Hence, it suffices to compare the conditional error $\Prob(\eventerror\given \valmaxdisadvantage > \valmaxadvantage, \eventbiasedreviewer)$ for $\eventerror\in \{\eventerrorseg, \eventerrorhol\}$. Let the random variable $\setapps\subseteq [\numapps]$ denote the set of disadvantaged applicants assigned to the unbiased reviewer. We now analyze the conditional error separately for part~\ref{item:attrs_one} and part~\ref{item:attrs_both}.

\subsection{Proof of Theorem~\ref{prop:compare_seg_hol_two_attrs}\ref{item:attrs_one}}

We analyze the conditional error $\Prob(\eventerror \given \valmaxdisadvantage>\valmaxadvantage, \eventbiasedreviewer)$ separately for holistic and segmented allocations.

\paragraph{Error for \segmented allocation.} 
Recall that in \segallo, each of the two reviewers is assigned one attribute. Since the reviewer assignment is independent from all else, by symmetry, the biased reviewer is assigned the protected attribute with probability $\frac{1}{2}$. In this case, the estimated quality of the best disadvantaged \applicant becomes $\frac{1 + \discount}{2}\valmaxdisadvantage$. On the other hand, with probability $\frac{1}{2}$ the unbiased reviewer is assigned the protected attribute. In this case, there is no discounting, and the best \applicant is always correctly identified. We have
\begin{subequations}\label{eq:proof_one_attr}
\begin{align}
    \Prob(\eventerrorseg\given \valmaxdisadvantage > \valmaxadvantage, \eventbiasedreviewer) = \frac{1}{2} \Prob\left(\frac{1+\discount}{2}\valmaxdisadvantage < \valmaxadvantage\given \valmaxdisadvantage > \valmaxadvantage, \eventbiasedreviewer\right).\label{proof_one_attr_seg_term}
\end{align}

\paragraph{Error for holistic allocation.}
Recall that in \holallo, each of the two reviewers is assigned both attributes of half of the \applicants. By symmetry, the biased reviewer is assigned the best disadvantaged \applicant with probability $\frac{1}{2}$. In this case, the estimated quality of the best disadvantaged \applicant is again $\frac{1+\discount}{2}\valmaxdisadvantage$. In order for this \applicant to be identified the best, its estimated quality needs to exceed both the best advantaged \applicant, and also all disadvantaged \applicants assigned to the unbiased reviewer who does not discount. 
On the other hand, with probability $\frac{1}{2}$ the unbiased reviewer is assigned the best disadvantaged \applicant. It is correctly identified as the best \applicant, because there is no discounting applied on this \applicant. 

Recall that random variable $\setapps\subseteq [\numapps]$ denote the set of disadvantaged applicants assigned to the unbiased reviewer. 
We have
\begin{align}
     \Prob(\eventerrorhol\given \valmaxdisadvantage>\valmaxadvantage, \eventbiasedreviewer) = \frac{1}{2} \Prob\left(\left\{\frac{1 + \discount}{2} \valmaxdisadvantage< \valmaxadvantage\right\} \union \left\{\frac{1+\discount}{2}\valmaxdisadvantage < \max_{\idxapp\in \setapps} \valdisadvantage_\idxapp\right\} \;\middle| \;\valmaxdisadvantage > \valmaxadvantage, \eventbiasedreviewer\right).\label{eq:proof_one_attr_hol_term}
\end{align}
\end{subequations}
\medskip
Combining~\eqref{eq:proof_one_attr}, we have
\begin{align*}
    \Prob(\eventerrorseg \given \valmaxdisadvantage>\valmaxadvantage, \eventbiasedreviewer)\le \Prob(\eventerrorhol\given \valmaxdisadvantage>\valmaxadvantage, \eventbiasedreviewer),
\end{align*}
completing the proof of part~\ref{item:attrs_one}.

\subsection{Proof of Theorem~\ref{prop:compare_seg_hol_two_attrs}\ref{item:attrs_both}}

We again analyze the conditional error $\Prob(\eventerror \given \valmaxdisadvantage>\valmaxadvantage, \eventbiasedreviewer)$ separately for holistic and segmented allocations.

\paragraph{Error for \segmented allocation.} 
When there is exactly one biased reviewer, one attribute of all disadvantaged \applicants is discounted, and the estimated quality of the best disadvantaged \applicant becomes $\frac{1 + \discount}{2}\valmaxdisadvantage$. In this case, the estimated quality of the best disadvantaged \applicant remains the best among the disadvantaged \applicants. It is correctly identified as the best \applicant, if and only if its estimated quality exceeds that of the best advantaged \applicant. We have
\begin{align}
    \Prob(\eventerrorseg \given \valmaxdisadvantage > \valmaxadvantage, \eventbiasedreviewer) & =\Prob\left(\frac{1+\discount}{2} \valmaxdisadvantage < \valmaxadvantage \;\middle|\; \valmaxdisadvantage > \valmaxadvantage, \eventbiasedreviewer\right).\label{eq:proof_both_attrs_seg}
\end{align}

\paragraph{Error for holistic allocation.} 
By symmetry, with probability $\frac{1}{2}$, the best disadvantaged applicant is assigned the unbiased reviewer. In this case, its estimated quality equals the true quality, and it is corrected identified as the best applicant. On the other hand, with probability $\frac{1}{2}$, the best disadvantaged applicant is assigned the biased reviewer. In this case, the estimated quality of this \applicant becomes $\discount\valmaxdisadvantage$, and it remains the best among all disadvantaged \applicants assigned to the biased reviewer. In order for this \applicant to be identified as the best \applicant, its estimated quality needs to exceed both the best advantaged \applicant, and also all disadvantaged \applicant assigned to the unbiased reviewer.
Recall that the random variable $\setapps\subseteq [\numapps]$ denote the set of disadvantaged applicants assigned to the unbiased reviewer. We have
\begin{align}
   \Prob(\eventerrorhol \given \valmaxdisadvantage > \valmaxadvantage, \eventbiasedreviewer) & = \frac{1}{2}\Prob\left(\left\{\discount\valmaxdisadvantage < \valmaxadvantage\right\} \union \left\{\discount \valmaxdisadvantage < \max_{\idxapp\in \setapps} \valdisadvantage_\idxapp\right\} \;\middle|\; \valmaxdisadvantage > \valmaxadvantage, \eventbiasedreviewer\right).\label{eq:proof_both_attrs_hol_term}
\end{align}
Now setting $\discount = 0$ in~\eqref{eq:proof_both_attrs_hol_term}, we have
\begin{align}
    \Prob(\eventerrorhol \given \valmaxdisadvantage > \valmaxadvantage, \eventbiasedreviewer) = \frac{1}{2}.\label{eq:proof_both_attrs_hol_case_one}
\end{align}

\paragraph{Comparing the error for segmented and holistic allocations.}

Subtracting~\eqref{eq:proof_both_attrs_seg} from~\eqref{eq:proof_both_attrs_hol_case_one}, we have that when $\discount=0$,
\begin{align}
    \Prob(\eventerrorhol \given \valmaxdisadvantage> \valmaxadvantage, \eventbiasedreviewer)  - \Prob(\eventerrorseg \given \valmaxdisadvantage> \valmaxadvantage, \eventbiasedreviewer) & = \frac{1}{2} - \Prob\left(\frac{1}{2} \valmaxdisadvantage < \valmaxadvantage \;\middle|\; \valmaxdisadvantage > \valmaxadvantage, \eventbiasedreviewer\right) \nonumber\\
    &= \Prob\left(\frac{1}{2} \valmaxdisadvantage > \valmaxadvantage \;\middle|\; \valmaxdisadvantage > \valmaxadvantage, \eventbiasedreviewer\right) -\frac{1}{2} \nonumber\\
    & \stackrel{\stepone}{=} \Prob\left(\frac{1}{2} \valmaxdisadvantage > \valmaxadvantage \;\middle|\; \valmaxdisadvantage > \valmaxadvantage\right) - \frac{1}{2} \nonumber\\
    & \stackrel{\steptwo}{=}  2 \cdot \Prob\left(\frac{1}{2} \valmaxdisadvantage > \valmaxadvantage \right) - \frac{1}{2} \label{eq:proof_both_attrs_subtract}
\end{align}
where step~\stepone is true because the quality values are independent of whether the reviewers are biased, and step~\steptwo is true because \begin{align*}
    \Prob\left(\frac{1}{2} \valmaxdisadvantage > \valmaxadvantage \;\middle|\; \valmaxdisadvantage > \valmaxadvantage\right)
=\frac{\Prob(\frac{1}{2} \valmaxdisadvantage > \valmaxadvantage, \valmaxdisadvantage > \valmaxadvantage)}{\Prob(\valmaxdisadvantage>\valmaxadvantage)}
= \frac{\Prob(\frac{1}{2} \valmaxdisadvantage > \valmaxadvantage)}{\Prob(\valmaxdisadvantage>\valmaxadvantage)} = 2\cdot \Prob\left(\frac{1}{2} \valmaxdisadvantage > \valmaxadvantage\right),
\end{align*}
where the last equality is true because $\Prob(\valmaxdisadvantage>\valmaxadvantage)=\frac{1}{2}$ by symmetry, when the fraction of disadvantaged \applicants is $\fracappsdisadvantage = 0.5$.

Hence, the difference in error between \segmented and \holistic allocations is
\begin{align}
    \Prob(\eventerrorhol) - \Prob(\eventerrorseg) & \stackrel{\stepone}{=} \frac{1}{2}\big[\Prob(\eventerrorhol\given \valmaxdisadvantage > \valmaxadvantage) - \Prob(\eventerrorseg\given \valmaxdisadvantage > \valmaxadvantage)\big] \nonumber\\
    & \stackrel{\steptwo}{=} \frac{\Prob(\eventbiasedreviewer)}{2}\cdot \big[\Prob(\eventerrorhol\given \valmaxdisadvantage > \valmaxadvantage,\eventbiasedreviewer) - \Prob(\eventerrorseg\given \valmaxdisadvantage > \valmaxadvantage,\eventbiasedreviewer)\big]\label{eq:proof_both_attrs_err}
\end{align}
where step~\stepone is true by~\eqref{eq:proof_both_attrs_err_reduce_disadvantage}, and step~\steptwo is true because the error of segmented and holistic allocations is identical conditional on $\setcomplement{\eventbiasedreviewer}$.
Plugging~\eqref{eq:proof_both_attrs_subtract} and the fact that $\Prob(\eventbiasedreviewer) = 2\fracbiased(1-\fracbiased)$ to~\eqref{eq:proof_both_attrs_err} and rearranging completes the proof of~\eqref{eq:prop_err_comparison} and~\eqref{eq:condition}.

\paragraph{Condition for power-law distribution.}
Following Definition 3 of~\citet{kleinberg2018rooney}, for non-negative functions $f(n)$ and $g(n)$, we define \begin{align*}
    f(n)\apeq g(n)
\end{align*}
if and only if $f(n) = g(n) \left(1\pm O\left(\frac{(\ln n)^2}{n}\right)\right)$. Now consider the power-law distribution with constant parameter $\parampowerlaw$. Setting $\alpha= 1, \beta = 2, c = \alpha \beta^{-(1+\delta)}$ and $k=1$ in Theorem B.3 of~\citet{kleinberg2018rooney} yields
\begin{align}
    \Prob(\valmaxdisadvantage <2 \valmaxadvantage) \apeq \left(1+2^{-(1+\delta)}\right)^{-1}.\label{eq:proof_both_attrs_prob_approx}
\end{align}
According to~\eqref{eq:condition}, \segallo is better than \holallo if and only if \begin{align*}
    \Prob(\valmaxdisadvantage >2 \valmaxadvantage) > 0.25,
\end{align*}
or equivalently
\begin{align}
    \Prob(\valmaxdisadvantage <2 \valmaxadvantage) < 0.75.\label{eq:condition_rewrite}
\end{align}
Combining~\eqref{eq:proof_both_attrs_prob_approx} and~\eqref{eq:condition_rewrite}, for sufficiently large $\numapps$, segmented allocation is better if and only if the constant $\parampowerlaw$ satisfies
\begin{align*}
    \left(1 + 2^{-(1+\parampowerlaw)}\right)^{-1} < 0.75,
\end{align*}
or equivalently $\parampowerlaw < \frac{\log 3}{\log 2} -1$, completing the proof of claim~\eqref{eq:condition_powerlaw}.

\section{Results from a previous experiment}\label{app:previous_expt}

In this section, we discuss a previous version of the experiment (termed ``previous experiment''), conducted before the new version presented in Section~\ref{sec:calibration}  (termed ``new experiment''). For completeness, we present results pertaining to the previous experiment. Based on these results, we also discuss our reason to collect data for the new experiment.

The setup of the previous experiment is identical to the new experiment described in Section~\ref{sec:calibration}, except for one difference in how the initial errors of the \grouptwenty (see the \grouptwentyinit curve in Figure~\ref{fig:calibration} and Figure~\ref{fig:calibration_old}) are computed, to be detailed later in this section. The main conclusions made in Section~\ref{sec:calibration} remain unchanged across these two experiments. We also describe a few other observations that are different across these two experiments, and discuss our conjectures of their causes.

\subsection{Results consistent with the new experiment} 
We report the results for the previous experiment, under the same data analysis procedure as described in Section~\ref{sec:calibration}. 
Comparing the \groupfive and the \grouptwenty, the workers' mean error in the previous experiment is $1.05 \pm 0.06$ in the \groupfive, and $ 0.69 \pm 0.04$ in the \grouptwenty. We perform a univariate permutation test, and reject the null hypothesis that the errors of the two groups have the same mean (one-sided $p$-value $<0.01$; Cohen's effect size $d= 0.70$).

Comparing the errors for individual pages within the \grouptwenty in the previous experiment, the results are shown in Figure~\ref{fig:calibration_old}. In the \grouptwenty, the (final) mean error for page $1$ is $0.76 \pm 0.05$, and the (final) mean error for page $4$ is $0.64 \pm 0.05$. We perform a univariate permutation test, and reject the null hypothesis that the mean errors for these two pages have the same mean (one-sided $p$-value $<0.01$; Cohen's effect size $d=0.25$).

The qualitative trends from these results are consistent with the results reported in the new experiment in Section~\ref{sec:calibration}.

\subsection{Observations different from the new experiment}
\begin{figure}[tb]
    \centering
    \includegraphics[width=0.35\linewidth]{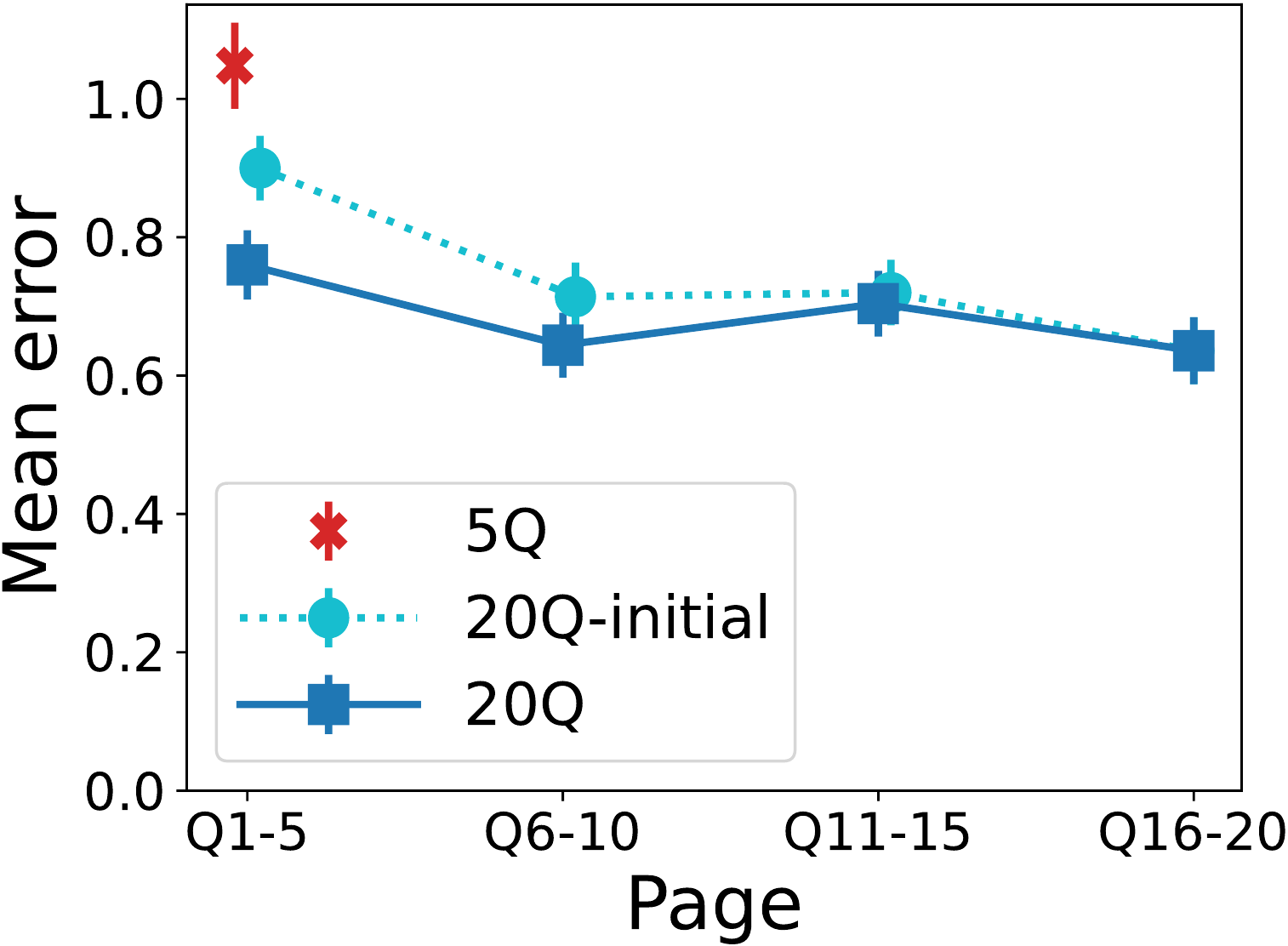}

    \caption{The mean error in estimating the percentile bins in the previous experiment, for workers in the \groupfive and the \grouptwenty. Error bars represent the standard error of the mean.}
    \label{fig:calibration_old}
\end{figure}

Recall that the \grouptwentyinit curve represents the initial mean error for each page, defined as the error of the worker's answers for each individual page, right before the worker ever turns to the next page. As described in Section~\ref{sec:calibration}, we expect that the initial mean error for page $1$ of the \grouptwenty is similar to the mean error of the \groupfive, because the two settings are identical before the workers in the \grouptwenty ever turns to page $2$. However, in Figure~\ref{fig:calibration_old} pertaining to the previous experiment, we observe that the initial mean error for page $1$ of the \grouptwenty ($0.90\pm 0.05$), as shown in the \grouptwentyinit curve, is notably smaller than the error in the \groupfive ($1.05 \pm 0.06$).
We make two hypotheses for this discrepancy in the previous experiment. 

\paragraph{Workers abandoning the task.}
We allow the workers to abandon the task at any time they wish, and in the data analysis only include the workers who have completed the task. Since the workers in the \grouptwenty answer more questions than the \groupfive, it is possible that more workers from the \grouptwenty abandon the task, and this self-selection leads to higher quality workers for the \grouptwenty. We re-compute the errors by including the answers from the abandoning workers back to the analysis, and observe negligible difference compared to when excluding these abandoning workers.
We thus determine that workers abandoning the task is not the primary cause of the discrepancy.

\paragraph{Inaccuracy in imputing page turning.}
In the previous experiment, we record the worker's answer to each question and its corresponding timestamp. More precisely, if a worker ever modifies their answer to a question, the initial answer and all subsequent modified answers are recorded along with their timestamps. In the previous experiment, timestamps directly associated with the action of turning pages are not recorded. As a remedy, we impute when the worker ever turns to the next page, by using the first timestamp that the worker ever answers a question on the next page. The initial mean error for the current page is then computed by using the answers right before this timestamp.

This imputation introduces inaccuracy, because it does not preclude the possibility where a worker turns to the next page, looks at the \applicants presented on the next page without answering any questions, and immediately turns back to the previous page to modify their answers. In this case, no answer to the next page is recorded, but the worker has indeed turned to the page, and then comes back to the previous page. It is thus possible that the worker uses information gained from the \applicants on the next page to modify answers on the current page, leading to the lower initial error for page 1 of the the \grouptwenty that the error of the \groupfive.\footnote{
    Recall that we allow the worker to turn back to previous pages at any time, but only turn to the next page if all questions on the current page have been answered, so the inaccuracy only comes from workers gaining information from the next one page, but not the subsequent pages later.
}

\paragraph{Change in experimental design.}
To understand how the imputation of page turning in the previous experiment contributes to the discrepancy between the error of the \groupfive and the initial error for page 1 of the \grouptwenty, we conduct the new experiment presented in Section~\ref{sec:calibration}. In the new experiment, the task is identical except that we additionally record the timestamps directly associated with each button click that turns pages. To compute the initial error for each individual page of the \grouptwenty, we now directly use the first timestamp that the worker ever clicks the button to turn to the next page, and compute the initial mean error of the current page using the worker's answers right before this timestamp. For comparison, in the new experiment, we also compute the errors using the previously imputation for page turning. We observe negligible difference in the \grouptwentyinit curve, when using the button timestamps versus using the page-turning imputation in the new experiment. We hence determine that the primary cause of the discrepancy between the \grouptwentyinit curve and the \groupfive in the previous experiment is not the page-turning imputation either. We do not have a clear alternative explanation for this discrepancy, and attribute it to randomness in data collection.

\paragraph{Difference in worker mean errors.} In addition to the discrepancy between the error of the \groupfive and the initial error for page 1 of the \grouptwenty in the previous experiment, we also observe that the errors are overall smaller in the previous experiment than in the new experiment. For example, the workers' mean error for the \groupfive and \grouptwenty are $1.05\pm 0.06$ and $0.69 \pm 0.04$ respectively in the previous experiment, compared to $1.14\pm 0.06$ and $0.84\pm 0.05$ respectively in the new experiment. We do not have a clear explanation for this difference. Since the two experiments only differ in whether the button click information is recorded, and this change is not visible to the workers, we conjecture that this difference in errors is related to worker quality changes specific to the crowdsourcing platform (e.g., different dates, day of the week, and time of the day) or randomness in data collection.

